%% file: main.tex
\pdfoutput=1
\documentclass[12pt,a4paper]{article}
\usepackage{ifthen} % for conditional statements
\newboolean{pdflatex}
\setboolean{pdflatex}{true} % False for eps figures 

\newboolean{articletitles}
\setboolean{articletitles}{true} % False removes titles in references

\newboolean{uprightparticles}
\setboolean{uprightparticles}{false} %True for upright particle symbols
\usepackage{siunitx}
\usepackage{booktabs}
\def\paperauthors{LHCb collaboration} % Leave as is for PAPER, CONF and FIGURE
\def\paperasciititle{Rapidity and multiplicity dependence of chargedparticle flow in pPb collisions at sqrt(s_NN) = 8.16 TeV} % Set ASCII title here !! MAKE sure it's only ASCII characters !! 

\def\papertitle{Rapidity and multiplicity dependence of charged-particle flow in $p$Pb collisions at $\sqsnn = 8.16\tev$} % Latex formatted title
\def\paperkeywords{{High Energy Physics}, {LHCb}} % Comma separated list
\def\papercopyright{\the\year\ CERN for the benefit of the LHCb collaboration} % new since 9/Apr/2018
\def\paperlicence{CC BY 4.0 licence}
\def\paperlicenceurl{https://creativecommons.org/licenses/by/4.0/}

\newif\ifEnableSectionTOCLinks
\EnableSectionTOCLinksfalse % deactivated

\input{preamble}
\usepackage{longtable} % only for template; not usually to be used in PAPERs

\newcommand{\deta}{{\ensuremath{\Delta \eta}}\xspace}
\newcommand{\dphi}{{\ensuremath{\Delta \phi}}\xspace}
\newcommand{\dndeta}{{\ensuremath{\deriv N_{\mathrm{ch}}/\deriv \eta}}\xspace}
\newcommand{\nch}{{\ensuremath{N_{\mathrm{ch}}}}\xspace}

\newcommand{\meandndeta}{{\ensuremath{\langle \deriv N_{\mathrm{ch}}/\deriv \eta \rangle}}\xspace}

\begin{document}

%%%%%%%%%%%%%%%%%%%%%%%%%
%%%%% Title     %%%%%%%%%
%%%%%%%%%%%%%%%%%%%%%%%%%
\renewcommand{\thefootnote}{\fnsymbol{footnote}}
\setcounter{footnote}{1}

% %%%%%%% CHOOSE TITLE PAGE--------
%\onecolumn
%\input{title-LHCb-INT}
%\input{title-LHCb-ANA}
%\input{title-LHCb-CONF}
%\input{title-LHCb-FIGURE}
\input{title-LHCb-PAPER}

%\twocolumn
% %%%%%%%%%%%%% ---------

\renewcommand{\thefootnote}{\arabic{footnote}}
\setcounter{footnote}{0}

%%%%%%%%%%%%%%%%%%%%%%%%%%%%%%%%
%%%%%  Table of Content   %%%%%%
%%%%%%%%%%%%%%%%%%%%%%%%%%%%%%%%
%%%% Uncomment if desired
%\tableofcontents

\cleardoublepage

%%%%%%%%%%%%%%%%%%%%%%%%%
%%%%% Main text %%%%%%%%%
%%%%%%%%%%%%%%%%%%%%%%%%%

\pagestyle{plain} % restore page numbers for the main text
\setcounter{page}{1}
\pagenumbering{arabic}

%% Uncomment during review phase. 
%% Comment before a final submission.
%\linenumbers

%% This is the main body
%% It is useful to have a single file so comments are not missed in overleaf.
{\input{body}}

% Do not include this in any draft (just for information in the template)
%\input{acknowledgements_intro}
% Comment this in for paper drafts; do not include this in analysis note, conference and figure reports
\input{acknowledgements}

\newpage

\input{appendix}

\clearpage

% This should be taken out in the final paper
%\input{supplementary-app}

\addcontentsline{toc}{section}{References}
%\setboolean{inbibliography}{true}
\bibliographystyle{LHCb}
%\bibliography{main,standard,LHCb-PAPER,LHCb-CONF,LHCb-DP,LHCb-TDR}
\input{main.bbl}

\newpage

\input{Authorship_LHCb-PAPER-2025-003}

\iffalse
The author list for journal publications is generated from the
Membership Database shortly after 'approval to go to paper' has been
given.  It is available at \url{https://lbfence.cern.ch/membership/authorship}
and will be sent to you by email shortly after a paper number
has been assigned.  
The author list should be included in the draft used for 
first and second circulation, to allow new members of the collaboration to verify
that they have been included correctly. Occasionally a misspelled
name is corrected, or associated institutions become full members.
Therefore an updated author list will be sent to you after the final
EB review of the paper.  In case line numbering doesn't work well
after including the authorlist, try moving the \verb!\bigskip! after
the last author to a separate line.

The authorship for Conference Reports should be ``The LHCb
collaboration'', with a footnote giving the name(s) of the contact
author(s), but without the full list of collaboration names.

The authorship for Figure Reports should be ``The LHCb
collaboration'', with no contact author and without the full list 
of collaboration names.
\fi

\end{document}

%% file: preamble.tex
\usepackage[top=1in, bottom=1.25in, left=1in, right=1in]{geometry}

\usepackage{microtype}
\usepackage{lineno}  % for line numbering during review
\usepackage{xspace} % To avoid problems with missing or double spaces after
\usepackage{caption} %these three command get the figure and table captions automatically small

\usepackage{graphicx}  % to include figures (can also use other packages)
\usepackage{color}
\usepackage{colortbl}
\graphicspath{{./figs/}} % Make Latex search fig subdir for figures
\usepackage{amsmath} % Adds a large collection of math symbols
\usepackage{amssymb}
\usepackage{amsfonts}
\usepackage{upgreek} % Adds in support for greek letters in roman typeset

\newcommand*\patchAmsMathEnvironmentForLineno[1]{%
\expandafter\let\csname old#1\expandafter\endcsname\csname #1\endcsname
\expandafter\let\csname oldend#1\expandafter\endcsname\csname
end#1\endcsname
 \renewenvironment{#1}%
   {\linenomath\csname old#1\endcsname}%
   {\csname oldend#1\endcsname\endlinenomath}%
}
\newcommand*\patchBothAmsMathEnvironmentsForLineno[1]{%
  \patchAmsMathEnvironmentForLineno{#1}%
  \patchAmsMathEnvironmentForLineno{#1*}%
}
\AtBeginDocument{%
\patchBothAmsMathEnvironmentsForLineno{equation}%
\patchBothAmsMathEnvironmentsForLineno{align}%
\patchBothAmsMathEnvironmentsForLineno{flalign}%
\patchBothAmsMathEnvironmentsForLineno{alignat}%
\patchBothAmsMathEnvironmentsForLineno{gather}%
\patchBothAmsMathEnvironmentsForLineno{multline}%
\patchBothAmsMathEnvironmentsForLineno{eqnarray}%
}

\usepackage[pdftex,
            pdfauthor={\paperauthors},
            pdftitle={\paperasciititle},
            pdfkeywords={\paperkeywords}]{hyperref}
\usepackage{hyperxmp}
\hypersetup{
    pdfcopyright={Copyright (C) \papercopyright},
    pdflicenseurl={\paperlicenceurl}
}
\usepackage[colorinlistoftodos,textsize=scriptsize]{todonotes}

\usepackage[bottom,flushmargin,hang,multiple]{footmisc}

\usepackage[all]{hypcap} % Internal hyperlinks to floats.

\input{lhcb-symbols-def} % Add in the predefined LHCb symbols

\hypersetup{
  colorlinks   = true, %Colours links instead of ugly boxes
  urlcolor     = blue, %Colour for external hyperlinks
  linkcolor    = blue, %Colour of internal links
  citecolor    = red   %Colour of citations
}

\ifEnableSectionTOCLinks
    \usepackage[explicit]{titlesec} % to change headings
    
    \let\oldcontentsline\contentsline
    \renewcommand

    \titleformat{\section}{\normalfont\Large\bf}{\hyperlink{tocsection.\thesection}{{\thesection} \parbox[t]{\dimexpr\textwidth-1pc}{#1}}}{1pc}{}

    \titleformat{\subsection}{\normalfont\bf}{\hyperlink{tocsubsection.\thesubsection}{{\thesubsection} \parbox[t]{\dimexpr\textwidth-1pc}{#1}}}{1pc}{}

    \titleformat{name=\section,numberless}[display]{}{}{0pt}{\normalfont\Huge\bfseries #1}
\fi

\usepackage{cite} % Allows for ranges in citations
\usepackage{mciteplus}

%% file: lhcb-symbols-def.tex
\usepackage{xspace} 
\usepackage{upgreek}

\def\lhcb   {\mbox{LHCb}\xspace}

\def\velo   {VELO\xspace}

\def\MagUp {\mbox{\em Mag\kern -0.05em Up}\xspace}

\ifthenelse{\boolean{uprightparticles}}%
{

 \def\PDelta      {\ensuremath{\Delta}\xspace}                 
 \def\PXi         {\ensuremath{\Xi}\xspace}                 
 \def\PLambda     {\ensuremath{\Lambda}\xspace}                 
 \def\PSigma      {\ensuremath{\Sigma}\xspace}                 
 \def\POmega      {\ensuremath{\Omega}\xspace}                 
 \def\PUpsilon    {\ensuremath{\Upsilon}\xspace}
 \let\oldPi\Pi
 \def\PPi         {\ensuremath{\oldPi}\xspace}

 \def\PB      {\ensuremath{\mathrm{B}}\xspace}                 
 \def\PD      {\ensuremath{\mathrm{D}}\xspace}                 

 \def\PK      {\ensuremath{\mathrm{K}}\xspace}                 
 \def\Ps      {\ensuremath{\mathrm{s}}\xspace}

 \def\thebaroffset{0.0em}
}
{

 \mathchardef\PDelta="7101
 \mathchardef\PXi="7104
 \mathchardef\PLambda="7103
 \mathchardef\PSigma="7106
 \mathchardef\POmega="710A
 \mathchardef\PUpsilon="7107
 \mathchardef\PPi="7105
 \def\PB      {\ensuremath{B}\xspace}                 
 \def\PD      {\ensuremath{D}\xspace}                 

 \def\PK      {\ensuremath{K}\xspace}                 
 \def\Ps      {\ensuremath{s}\xspace}

 \def\thebaroffset{0.18em}
}
\newcommand{\offsetoverline}[2][\thebaroffset]{\kern #1\overline{\kern -#1 #2}}%

\makeatletter
\ifcase \@ptsize \relax% 10pt
  \newcommand{\miniscule}{\@setfontsize\miniscule{4}{5}}% \tiny: 5/6
\or% 11pt
  \newcommand{\miniscule}{\@setfontsize\miniscule{5}{6}}% \tiny: 6/7
\or% 12pt
  \newcommand{\miniscule}{\@setfontsize\miniscule{5}{6}}% \tiny: 6/7
\fi
\makeatother

\DeclareRobustCommand{\optbar}[1]{\shortstack{{\miniscule (\rule[.5ex]{1.25em}{.18mm})}
  \\ [-.7ex] $#1$}}

\def\squark    {{\ensuremath{\Ps}}\xspace}

\def\KorKbar {\kern \thebaroffset\optbar{\kern -\thebaroffset \PK}{}\xspace}

\def\D       {{\ensuremath{\PD}}\xspace}

\def\DorDbar {\kern \thebaroffset\optbar{\kern -\thebaroffset \PD}\xspace}

\def\Dp      {{\ensuremath{\D^+}}\xspace}
\def\Dm      {{\ensuremath{\D^-}}\xspace}

\def\DpDm    {\ensuremath{\Dp {\kern -0.16em \Dm}}\xspace}

\def\B       {{\ensuremath{\PB}}\xspace}

\def\BorBbar {\kern \thebaroffset\optbar{\kern -\thebaroffset \PB}\xspace}

\def\Bd      {{\ensuremath{\B^0}}\xspace}

\def\BdorBdbar {\kern \thebaroffset\optbar{\kern -\thebaroffset \Bd}\xspace}

\def\Bs      {{\ensuremath{\B^0_\squark}}\xspace}

\def\BsorBsbar {\kern \thebaroffset\optbar{\kern -\thebaroffset \Bs}\xspace}

\def\Y#1S{\ensuremath{\PUpsilon{(#1S)}}\xspace}

\def\LorLbar     {\kern \thebaroffset\optbar{\kern -\thebaroffset \PLambda}\xspace}

\def\AT#1     {\ensuremath{A_{\mathrm{T}}^{#1}}\xspace}           % 2

\def\C#1      {\ensuremath{\mathcal{C}_{#1}}\xspace}                       % 9
\def\Cp#1     {\ensuremath{\mathcal{C}_{#1}^{'}}\xspace}                    % 7
\def\Ceff#1   {\ensuremath{\mathcal{C}_{#1}^{\mathrm{(eff)}}}\xspace}        % 9  
\def\Cpeff#1  {\ensuremath{\mathcal{C}_{#1}^{'\mathrm{(eff)}}}\xspace}       % 7
\def\Ope#1    {\ensuremath{\mathcal{O}_{#1}}\xspace}                       % 2
\def\Opep#1   {\ensuremath{\mathcal{O}_{#1}^{'}}\xspace}                    % 7

\newcommand{\nospaceunit}[1]{\ensuremath{\text{#1}}}       
\newcommand{\aunit}[1]{\ensuremath{\text{\,#1}}}       
\newcommand{\tev}{\aunit{Te\kern -0.1em V}\xspace}
\newcommand{\gev}{\aunit{Ge\kern -0.1em V}\xspace}
\newcommand{\mev}{\aunit{Me\kern -0.1em V}\xspace}
\newcommand{\kev}{\aunit{ke\kern -0.1em V}\xspace}
\newcommand{\ev}{\aunit{e\kern -0.1em V}\xspace}
 
\newcommand{\mevc}{\ensuremath{\aunit{Me\kern -0.1em V\!/}c}\xspace}
\newcommand{\gevc}{\ensuremath{\aunit{Ge\kern -0.1em V\!/}c}\xspace}
\newcommand{\mevcc}{\ensuremath{\aunit{Me\kern -0.1em V\!/}c^2}\xspace}
\newcommand{\gevcc}{\ensuremath{\aunit{Ge\kern -0.1em V\!/}c^2}\xspace}
\def\cm   {\aunit{cm}\xspace}

\def\mm   {\aunit{mm}\xspace}

\def\mum  {\ensuremath{\,\upmu\nospaceunit{m}}\xspace}

\def\deriv {\ensuremath{\mathrm{d}}}

\def\gsim{{~\raise.15em\hbox{$>$}\kern-.85em
          \lower.35em\hbox{$\sim$}~}\xspace}
\def\lsim{{~\raise.15em\hbox{$<$}\kern-.85em
          \lower.35em\hbox{$\sim$}~}\xspace}

\def\sqs   {\ensuremath{\protect\sqrt{s}}\xspace}
\def\sqsnn {\ensuremath{\protect\sqrt{s_{\scriptscriptstyle\text{NN}}}}\xspace}
\def\pt         {\ensuremath{p_{\mathrm{T}}}\xspace}

\def\ptot       {\ensuremath{p}\xspace}

\def\evtgen     {\mbox{\textsc{EvtGen}}\xspace}

\def\geant      {\mbox{\textsc{Geant4}}\xspace}

\def\photos     {\mbox{\textsc{Photos}}\xspace}

\def\tell1  {TELL1\xspace}
\def\ukl1   {UKL1\xspace}

\newcommand{\ie}{\mbox{\itshape i.e.}\xspace}

\newcommand{\lhcborcid}[1]{\href{https://orcid.org/#1}{\hspace*{0.1em}\raisebox{-0.45ex}{\includegraphics[width=1em]{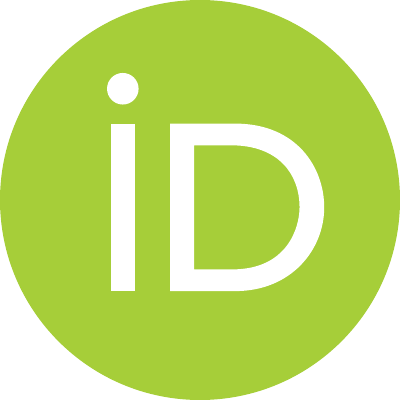}}}}

%% file: title-LHCb-PAPER.tex
% ===============================================================================
% Purpose: LHCb-PAPER journal paper title page template
% Author: 
% Created on: 2010-09-25
% ===============================================================================

%%%%%%%%%%%%%%%%%%%%%%%%%
%%%%%  TITLE PAGE  %%%%%%
%%%%%%%%%%%%%%%%%%%%%%%%%
\begin{titlepage}
\pagenumbering{roman}

% Header ---------------------------------------------------
\vspace*{-1.5cm}
\centerline{\large EUROPEAN ORGANIZATION FOR NUCLEAR RESEARCH (CERN)}
\vspace*{1.5cm}
\noindent
\begin{tabular*}{\linewidth}{lc@{\extracolsep{\fill}}r@{\extracolsep{0pt}}}
\ifthenelse{\boolean{pdflatex}}% Logo format choice
{\vspace*{-1.5cm}\mbox{\!\!\!\includegraphics[width=.14\textwidth]{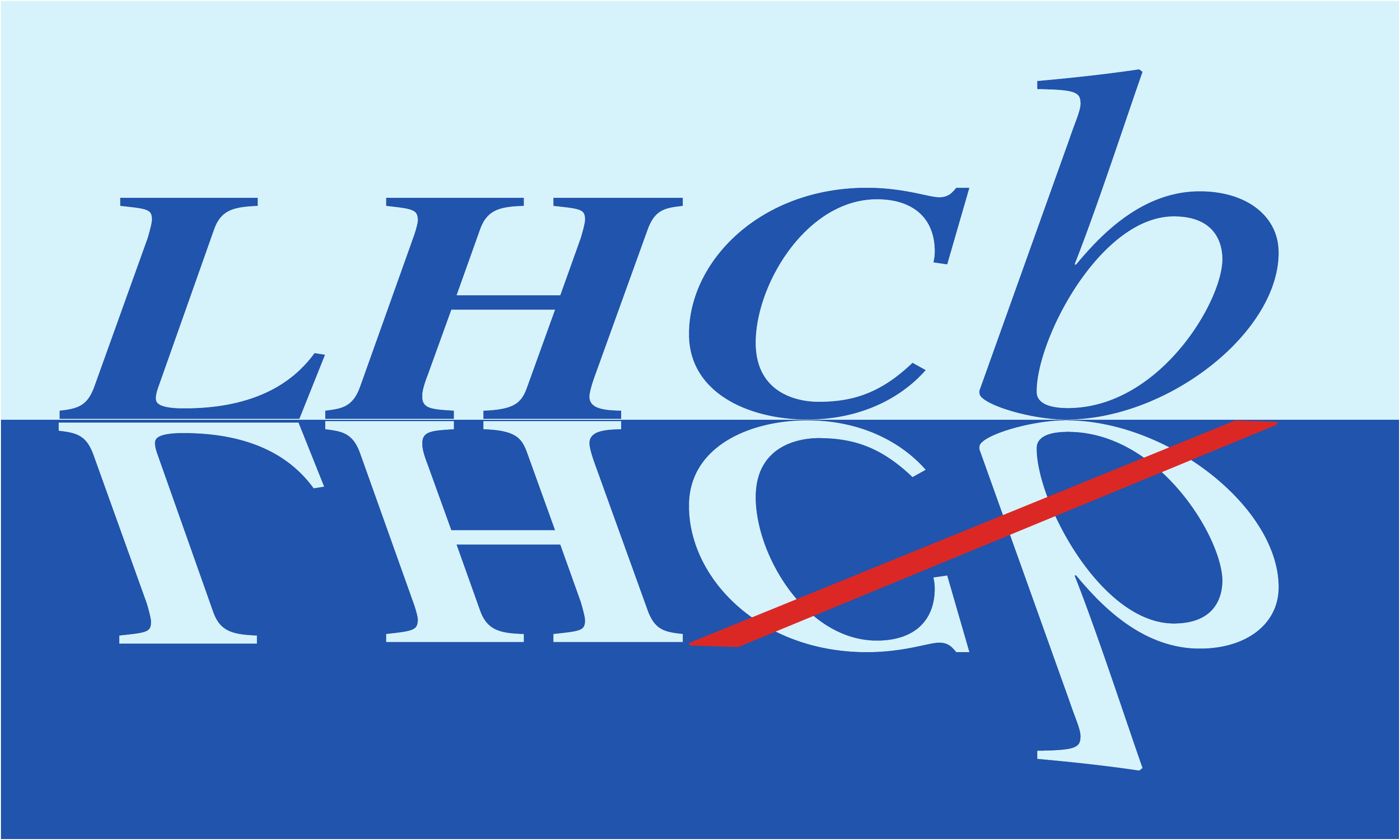}} & &}%
{\vspace*{-1.2cm}\mbox{\!\!\!\includegraphics[width=.12\textwidth]{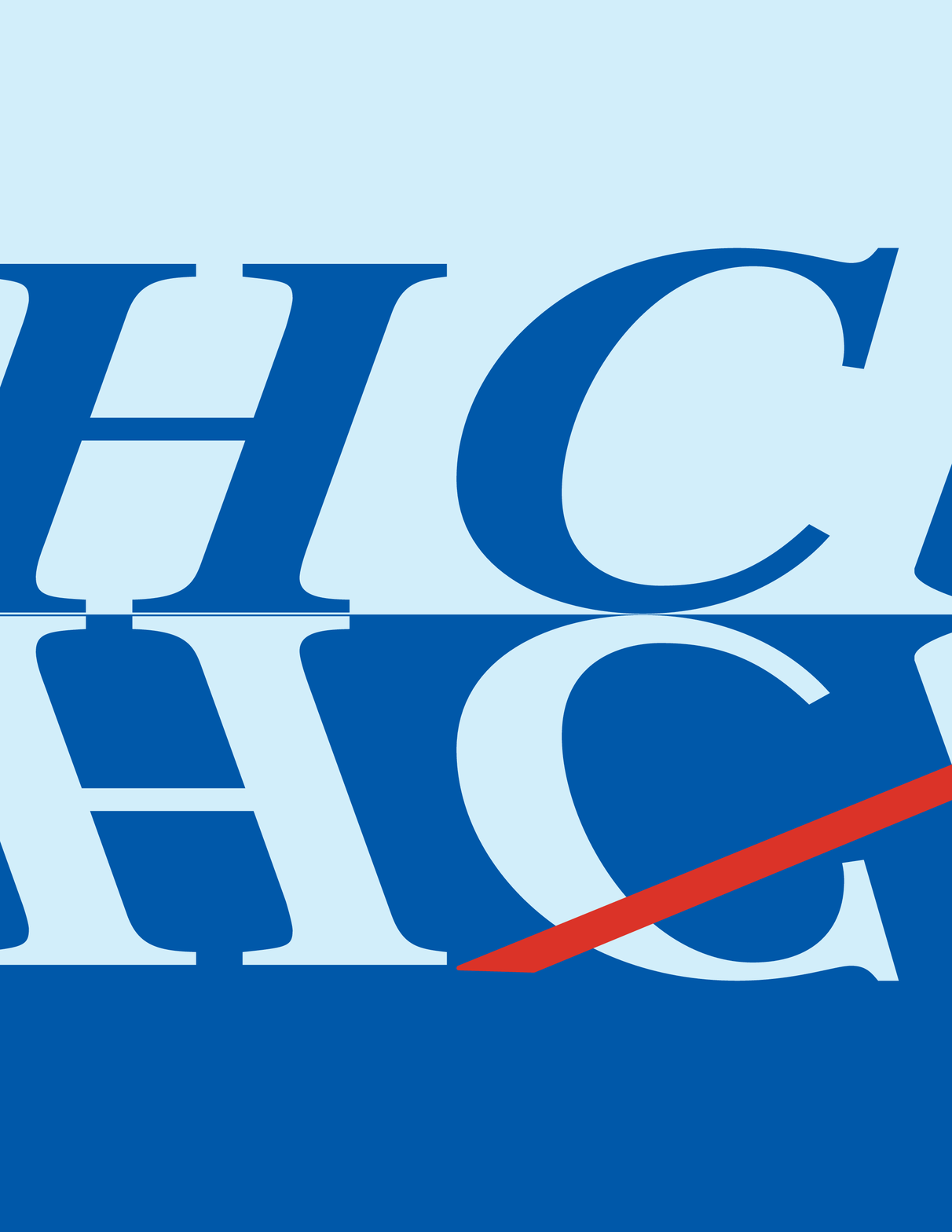}} & &}%
\\
 & & CERN-EP-2025-090 \\  % ID 
 & & LHCb-PAPER-2025-003 \\  % ID 
 & & 30 October 2025 \\       % \today \\ % Date - Can also hardwire e.g.: 23 March 2010
 & & \\
% not in paper \hline
\end{tabular*}

\vspace*{4.0cm}

% Title --------------------------------------------------
{\normalfont\bfseries\boldmath\huge
\begin{center}
% DO NOT EDIT HERE. Instead edit macro in main.tex to keep metadata correct
  \papertitle 
\end{center}
}

\vspace*{2.0cm}

% Authors -------------------------------------------------
\begin{center}

\paperauthors\footnote{Authors are listed at the end of this paper.}
\end{center}

\vspace{\fill}

% Abstract -----------------------------------------------
\begin{abstract}
  \noindent
 The elliptic and triangular flow of charged particles are measured using two-particle angular correlations in $p$Pb collisions in the pseudorapidity range \mbox{2.0 $< |\eta| <$ 4.8}. The data sample was collected by the LHCb experiment in 2016 at a centre-of-mass energy per nucleon pair of $\sqrt{s_{\rm NN}} = 8.16$ TeV, 
 containing in total approximately 1.5 billion collision events.
 Non-flow contributions are obtained in low-multiplicity collisions and subtracted to extract the flow harmonics. The results are presented as a function of event multiplicity and hadron transverse momentum. Comparisons with a full (3+1)D dynamic model indicate that it overestimates the measured elliptic flow. A comparison between the forward and backward regions reveals no significant differences in flow parameters, suggesting that final-state effects may dominate over initial-state effects in the origin of flow in small systems.
    
\end{abstract}

\vspace*{2.0cm}

\begin{center}
  Published in JHEP 10 (2025) 124
\end{center}

\vspace{\fill}

{\footnotesize 
\centerline{\copyright~\papercopyright. \href{\paperlicenceurl}{\paperlicence}.}}
\vspace*{2mm}

\end{titlepage}

%%%%%%%%%%%%%%%%%%%%%%%%%%%%%%%%
%%%%%  EOD OF TITLE PAGE  %%%%%%
%%%%%%%%%%%%%%%%%%%%%%%%%%%%%%%%

%  empty page follows the title page ----
\newpage
\setcounter{page}{2}
\mbox{~}

%% file: body.tex
\section{Introduction}
\label{sec:Introduction}

In ultrarelativistic heavy-ion collisions, strongly interacting QCD matter known as the Quark-Gluon Plasma (QGP) can be formed, 
in which deconfined quarks and gluons constitute a nearly ideal fluid~\cite{ALICE:2022wpn,STAR:2005gfr,PHENIX:2004vcz,Busza:2018rrf, hydro_review,Bernhard:2019bmu}. Extensive investigations have been conducted to understand and characterise this exotic form of matter in past and ongoing experiments at the relativistic heavy-ion collider (RHIC) and the LHC.  
Among the various strategies that have been developed over the years to identify signatures of the deconfined matter, flow analyses~\cite{Foka:2016vta} have emerged as an important approach to study the thermodynamic properties of the QGP. Since the QGP behaves akin to a quasi-ideal fluid, its evolution can be described using hydrodynamic equations~\cite{Ollitrault_2008}. In a semicentral collision of two nuclei AA at RHIC or the LHC, the overlap of the two colliding nuclei has an almond-shaped initial geometry. The pressure gradients on the short axis and the long axis differ. These differences induce a momentum-space anisotropy which can be measured as a modulation of the azimuthal distribution of the produced particles.
The phenomenon is known as collective flow and can be effectively described by hydrodynamic models~\cite{PhysRevD.46.229, OLLITRAULT201375c}.
This modulated particle azimuthal angular distribution can be described using a Fourier expansion, where the second harmonic coefficient is referred to as the elliptic flow ($v_2$) and the third harmonic coefficient as the triangular flow ($v_3$). Significant flow coefficients, even at higher orders, have been observed in many measurements at the LHC and RHIC~\cite{NA49:2003njx,STAR:2000ekf,STAR:2002hbo,STAR:2005npq,PHENIX:2003qra,PHENIX:2006dpn,PHOBOS:2010ekr,ALICE:2011ab,ALICE:2013xri,ALICE:2015cjr,ALICE:2016ccg,ALICE:2022zks,ATLAS:2011ah}. These coefficients are directly related to the hydrodynamic response of the created matter to the initial collision geometry. 
For example, by measuring the anisotropic particle flow in PbPb collisions, it becomes possible to estimate hydrodynamic parameters such as the shear viscosity of the QGP~\cite{Plumari:2019gwq}.

In $pp$ or $p$A collision systems, referred to as ``small systems",
the mean free path of produced partons is large enough that they propagate and fragment without significant interactions with each other~\cite{Grosse-Oetringhaus:2024bwr}. Thus, the presence of a final-state anisotropy in those systems was not anticipated.  
However, 
a raised long and narrow (``ridge-like") structure,
characterised by a similar angular modulation
as produced in AA collisions, 
has been observed in high-multiplicity (\ie, a large number of produced particles) $pp$ and $p$A collisions at the LHCb, ALICE, ATLAS, and CMS experiments at the LHC~\cite{LHCb-PAPER-2015-040,CMS:2012qk,ATLAS:2012cix,ALICE:2012eyl} as well as at the STAR and PHENIX experiments at RHIC~\cite{PhysRevLett.120.062302,2018,PhysRevLett.130.242301}. 
Recent efforts have been devoted to searching for the smallest system in which anisotropic flow can develop. 
Measurements of two-particle angular correlations, which serve as a probe of collective flow, are performed in collision systems with even smaller interaction region and different initial states. 
Studies in electron-positron ($e^+e^-$)~\cite{PhysRevLett.123.212002,CHEN2024138957, PhysRevLett.128.142005, Chen2023}, electron-proton ($ep$)~\cite{Abt2020,Abt2021} and photon-proton ($\gamma p$)~\cite{2023137905} collisions find no conclusive evidence of collective behaviour, however, nonzero values of the second- and third-order flow coefficients are reported in photon-lead ($\gamma$Pb) collisions\cite{PhysRevC.104.014903}. 
More recently, enhanced long-range elliptic anisotropies are also observed inside high-multiplicity jets produced in 13\tev $pp$ collisions~\cite{PhysRevLett.133.142301}.
While the exact origin of the anisotropic flow is still an area of active research and debate, there are several proposed explanations for this observation including hydrodynamic-like behaviour in these smaller collision systems that could exhibit a small-scale version of collective flow~\cite{Song2017}. Alternatively, initial-state effects such as gluon saturation~\cite{PhysRevLett.108.252301} could generate long-range correlations that manifest as the ridge in the final-state particle distributions. Other scenarios include final-state correlations that induce collective behaviour through multiple scattering~\cite{BASS1998255,Tarasovicova:2024isp}. Colour reconnection by which partons' colour connections rearrange during the fragmentation process can also lead to a ridge-like structure~\cite{PhysRevLett.111.042001}.

The \lhcb experiment enables the investigation of initial-state effects in high-energy collisions. 
Thanks to its unique forward experimental setup, the \lhcb detector geometry allows the study of two  centre-of-mass rapidity ($y^*$) regions of $p$Pb collisions: forward ($p$Pb, $1.5 < y^* < 4.0$) and backward (Pb$p$, $-5.0 < y^* < -2.5$). 
Forward collisions in LHCb provide access to the low-$x$ Bjorken region of the lead nucleus and measurements can be used to constrain nuclear parton distribution functions (nPDFs), where effects related to potential gluon saturation are anticipated. On the other hand, backward collisions are sensitive to higher $x$ values in the nPDFs. 
This differential sensitivity in the forward and backward collisions enables a comprehensive exploration of the diverse kinematic regions. 
Comparison of anisotropic flow in the two configurations sheds light on the role of small and intermediate-$x$ physics on the flow development and helps to understand its origin.

This paper reports the first measurement of charged-particle flow harmonic coefficients in $p$Pb collisions at \lhcb using two-particle angular correlation functions in the pseudorapidity region \mbox{$2.0<|\eta|<4.8$}. The elliptic and triangular flow harmonic coefficients $v_2$ and $v_3$ are extracted as functions of transverse momentum (\pt), and charged track density, $\dndeta$.
These studies complement the existing measurements by other experiments in different rapidity intervals covering different initial-state effects. The results are compared to theoretical calculations using a full $(3+1)$D dynamical framework with hydrodynamics and hadronic transport\cite{PhysRevLett.129.252302, PhysRevC.105.064905}.

%################################################
\section{Experimental setup}
\label{sec:LHCb}

The \lhcb detector~\cite{LHCb-DP-2008-001,LHCb-DP-2014-002} is a single-arm forward spectrometer covering the \mbox{pseudorapidity} range \mbox{$2<\eta <5$}. 
The detector during the LHC Run 2 period includes a high-precision tracking system consisting of a silicon-strip vertex detector (\velo)~\cite{LHCb-DP-2014-001} surrounding the interaction region, a large-area silicon-strip detector located upstream of a dipole magnet with a bending power of about $4{\mathrm{\,T\,m}}$, and three stations of silicon-strip detectors and straw drift tubes placed downstream of the magnet~\cite{LHCb-DP-2013-003,LHCb-DP-2017-001}.
The tracking system provides a measurement of the momentum, \ptot, of charged particles with a relative uncertainty that varies from 0.5\% at low momentum to 1.0\% at 200\gevc.
The minimum distance of a track to a primary collision vertex (PV), the impact parameter, is measured with a resolution of \mbox{$(15+29/\pt)\mum$}, where \pt is measured in\,\gevc.
Different types of charged hadrons are distinguished using information from two ring-imaging Cherenkov detectors. Photons, electrons and hadrons are identified by a calorimeter system. Muons are identified by a system composed of alternating layers of iron and multiwire proportional chambers. 

The online event selection is performed by a trigger. It consists of a hardware stage, which in this analysis randomly selects a predefined fraction of all bunch crossings,
followed by a software stage which requires at least one reconstructed track in the \velo.

The analysis is based on the proton and ${}^{208}$Pb ion collision data recorded by \lhcb during the LHC heavy-ion run in 2016. The centre-of-mass energy per nucleon pair is \mbox{$\sqsnn = 8.16\tev$}. The data were collected in two beam configurations. In the forward configuration, denoted $p$Pb, the proton beam travels in the direction from the \velo to the tracking system. 
In the backward configuration, denoted Pb$p$, the directions of the beams are reversed. 
After selection the forward and backward data samples contain $0.67$ and $0.83$ billion events, respectively.  

The flow effects are measured using prompt charged particles, 
which can be hadrons or leptons, with a mean lifetime 
\mbox{$\tau > 30$ ps}, 
produced directly in the collision or from decays of shorter-lifetime particles. They include all the particles detectable by the tracking system of LHCb (predominately $\pi^-$, $K^-$, $p$, $e^-$, $\mu^-$ and their antiparticles) and several hyperons which possess a net strangeness content~\cite{ALICE-PUBLIC-2017-005}. This translates in the analysis to the use of tracks that come directly from the primary vertex, with the selection discussed in Sec.~\ref{sec:data}.

Simulation is used to study the detector performance for prompt charged particles.
In the simulation, $p$Pb and Pb$p$ collisions are generated using the 
EPOS-LHC event generator~\cite{PhysRevC.92.034906}.
Decays of unstable particles are described by \evtgen~\cite{Lange:2001uf}, in which final-state radiation is generated using \photos~\cite{davidson2015photos}.
The interaction of the generated particles with the detector, and its response, are implemented using the \geant toolkit~\cite{Allison:2006ve, *Agostinelli:2002hh} as described in Ref.~\cite{LHCb-PROC-2011-006}.

%################################################
\section{Data selection}
\label{sec:data}

Only events containing fewer than 8000 \velo clusters are recorded for use in this analysis. This causes the loss of the $0.1\%$ highest-multiplicity events in the Pb$p$ sample while the fraction is negligible in the $p$Pb sample. Each collision event is required to contain exactly one reconstructed primary vertex, which is composed of at least five \velo tracks~\cite{Kucharczyk:1756296}. The position of the PV must be within $\pm 3$ standard deviations of the mean interaction position along the beam axis. The mean and the standard deviation of the luminous region are estimated from a Gaussian fit to the distributions of the PV position separately for the $p$Pb and Pb$p$ samples. The standard deviations are found to be around $44\mm$ for both datasets.

In the measurement, charged particles are reconstructed from tracks that hit all the subdetectors in the tracking system (so-called long tracks). 
Tracks are required to lie within the kinematic region defined by \mbox{$2.0 < \eta < 4.8$}, \mbox{$p>2.0\gevc$}, and \mbox{$0.2<\pt<5.0\gevc$}.
Fake tracks,\footnote{Fake tracks are those reconstructed from unrelated hits, which do not correspond to a genuine particle trajectory.} which are reconstruction artifacts, are suppressed using a multivariate classifier~\cite{DeCian:2255039}.
The impact parameter of the track with respect to the PV must be less than $1.0\mm$ to select tracks originating directly from the collision.

Simulation is used to study the reconstruction and selection efficiency due to the selection cuts, as well as the fake track and the secondary track contamination rates in the sample. These quantities are estimated as functions of the azimuthal angle $\phi$ and pseudorapidity, the transverse momentum, and an event multiplicity variable $N_{\rm \velo}^{\rm hits}$, which denotes the number of hits in the \velo for a given $p$Pb or Pb$p$ collision. A weight is calculated for each track in the sample
\begin{equation}
    w(\phi, \eta, \pt, N_{\rm \velo}^{\rm hits}) = (1-P_{\rm fake} - P_{\rm sec})/\epsilon ,
\end{equation}
where $P_{\rm fake}$ and $P_{\rm sec}$ are the contamination rates of fake and secondary tracks in the sample, and $\epsilon$ is the combined reconstruction and selection efficiency. The weighted sum of all the selected tracks in a given collision event is denoted as \nch, which stands for the number of produced charged particles in the kinematic region. 
To estimate the fully corrected charged-particle density without momentum requirements, a momentum extrapolation based on the EPOS-LHC simulation is performed. The charged-particle density, \dndeta, is obtained by applying a scaling factor
\begin{equation}
    \frac{\deriv N_{\rm ch}}{\deriv\eta} = \frac{N_{\rm ch}}{\varepsilon_{\rm ch} \Delta\eta},
\end{equation}
\noindent where $\varepsilon_{\rm ch}$ is the fraction of charged particles from simulation which 
lie within the accepted momentum region
and varies from $0.662$ to $0.688$ depending on $\nch$, and \mbox{$\Delta\eta=2.8$} is the pseudorapidity range.

Five event-multiplicity classes are defined based on $\nch$, where each contains around 10\% of the total collision events in the specific configuration, except for the lowest multiplicity class in the Pb$p$ sample which contains around 20\%. 
In total, the top 50--60\% events in multiplicity are included in the measurement.
The corresponding mean charged particle densities \meandndeta of the classes are estimated and are used to denote the classes. 
The \meandndeta uncertainties are described at the end of Sec.~\ref{sec:systematics}.
A summary of the event-multiplicity class definitions is shown in Table~\ref{tab:nch_bins}.
In addition to these five classes, a low-multiplicity class is also defined for each data configuration. This class is used  solely for the subtraction of non-flow effects, which is described in Sec.~\ref{sec:extract_vn}.

%################################################
\begin{table}[!t]
  \caption{
    \small 
    Event-multiplicity class definitions based on $\nch$ and the corresponding mean charged particle density $\meandndeta$ for $p$Pb and Pb$p$ configurations. The \meandndeta uncertainties are described at the end of Sec.~\ref{sec:systematics}. }
    \begin{center}\begin{tabular}{c c c}
    \hline
       &   $\nch$ class range    & $\meandndeta$    \\ 
    \hline
    $p$Pb 
           & $26 < \nch \le 31$      & $15.6 \pm 0.5$ \\
           & $31 < \nch \le 37$      & $18.2 \pm 0.6$\\
           & $37 < \nch \le 44$      & $21.8 \pm 0.8$\\
           & $44 < \nch \le 54$      & $25.9 \pm 0.9$\\
           & $\phantom{00000}\nch > 54$   & $34.5 \pm 1.2$ \\
    \hline
    Pb$p$  
           & $24 < \nch \le 39$      & $17.1 \pm 0.7$ \\
           & $39 < \nch \le 48$      & $23.4 \pm 0.9$ \\
           & $48 < \nch \le 59$      & $28.4 \pm 1.1$ \\
           & $59 < \nch \le 75$      & $35.0 \pm 1.4$ \\
           & $\phantom{00000}\nch > 75$             & $47.8 \pm 1.9$ \\
    \hline
  \end{tabular}\end{center}
\label{tab:nch_bins}
\end{table}
%################################################

\section{Two-particle correlation analysis}
\label{sec:2PC}

Angular correlation functions of pairs of charged particles, regardless of their charge sign, are constructed separately for each event-multiplicity class. 
Two particles 
in the same event, 
referred to as the trigger and associated particles,
are paired to form the angular distribution $S(\deta, \dphi)$, where $\deta$ and $\dphi$ are the differences in $\eta$ and $\phi$ between the pair. 
Both trigger and associated particles are from the selected prompt charged particle sample, but they are distinguished by separate \pt requirements, which are described later.
The mixed event angular distribution, $B(\deta, \dphi)$, is used to correct for detector effects, which mainly originate from the limited detector acceptance. To construct $B(\deta, \dphi)$, all trigger particles from an event are paired with all associated particles in ten different random events in the same event-multiplicity class, whose PV positions along the $z$ direction are within $2.2\cm$ of the original event. The two-particle angular correlation function is defined by
\begin{equation}
    C(\deta, \dphi) \equiv \frac{1}{N_{\rm trig}}\frac{\deriv^2N_{\text{pair}}}{\deriv\Delta\eta \, \deriv\Delta\phi} = B(0,0)  \frac{S(\Delta\eta, \Delta\phi)}{B(\Delta\eta, \Delta\phi)} ,
    \label{eq:cf}
\end{equation}
where $N_{\text{pair}}$ is the number of charged particle pairs in the (\deta, \dphi) bin, $N_{\rm trig}$ is the number of trigger particles within the trigger $\pt$ selection and $B(0,0)$ is a normalisation factor such that the detector pair acceptance is unity for \mbox{$\deta=0$, $\dphi=0$}, which is expected for two particles travelling along the same direction. 
The one-dimensional azimuthal correlation function, $C(\dphi)$, is obtained similarly, by projecting $S(\deta, \dphi)$ and $B(\deta, \dphi)$ onto the $\dphi$ axis in a common selected $\deta$ region and taking the ratio between the two. 

The trigger and associated particles are selected with independent $\pt$ requirements, which can be the same or different. The minimum \pt value for both trigger and associated particles is $0.8\gevc$ to ensure a uniform $\eta$ acceptance. 
In this work, the trigger and associated particles are selected with the same \pt requirement \mbox{$0.8 < \pt <5.0 \gevc$}. In addition to the full \pt range, the trigger particles are split in 10 smaller \pt intervals in order to study the \pt dependence of flow.

\begin{figure}[tb]
    \begin{center}
        \includegraphics[width=1.0\linewidth]{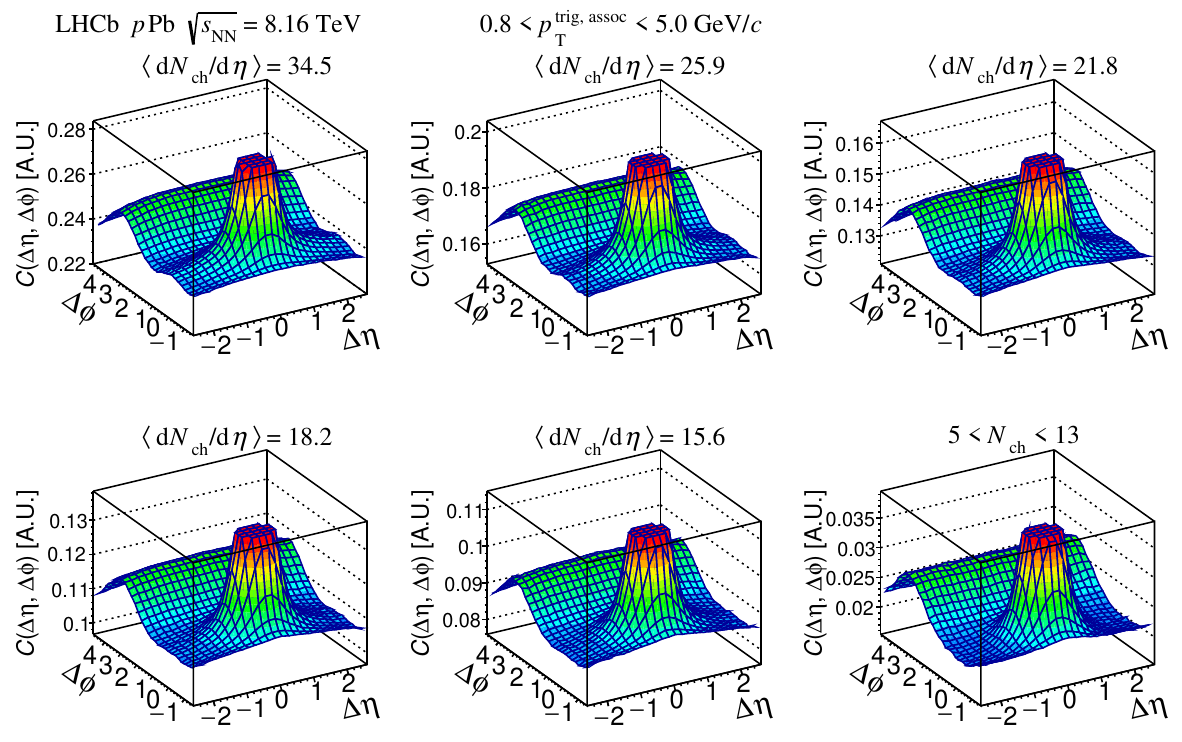}
        \vspace*{-0.5cm}
    \end{center}
    \caption{\small Example angular correlation functions $C(\Delta\eta, \Delta\phi)$ for the five multiplicity classes and the low-multiplicity class in $p$Pb collisions. The correlation functions are truncated at the peaks to improve the visibility of the ridge structures.
    }
    \label{fig:cf2d_pPb}
\end{figure}

\begin{figure}[tb]
    \begin{center}
        \includegraphics[width=1.0\linewidth]{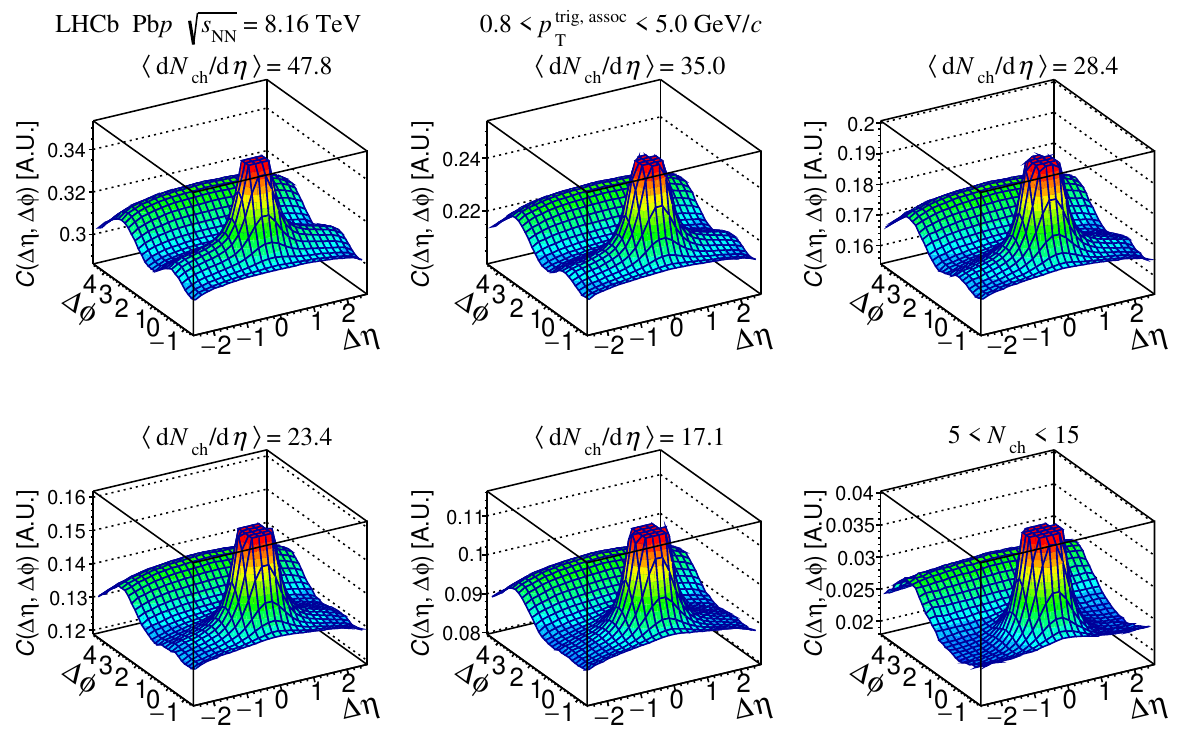}
        \vspace*{-0.5cm}
    \end{center}
    \caption{\small Example angular correlation functions $C(\Delta\eta, \Delta\phi)$ for the five multiplicity classes and the low-multiplicity class in Pb$p$ collisions. The correlation functions are truncated at the peaks to improve the visibility of the ridge structures. 
    }
    \label{fig:cf2d_Pbp}
\end{figure}

Figures~\ref{fig:cf2d_pPb} and~\ref{fig:cf2d_Pbp} show examples of the angular correlation functions $C(\deta, \dphi)$ in $p$Pb and Pb$p$ data. 
The correlation functions are presented in the five multiplicity classes defined in Table~\ref{tab:nch_bins} and in the low-multiplicity class described later in Sec.~\ref{sec:extract_vn}. 
A prominent peak at \mbox{$(\deta \simeq 0, \dphi \simeq 0)$}, which results from short-range non-flow interactions such as jets and particle decays, is observed in all the correlation functions. 
The peak is shown truncated to improve visibility of the lower structures in the correlation functions.
Ridge structures at \mbox{$\dphi \simeq 0$} (near-side) and \mbox{$\dphi \simeq \pi$} (away-side) for the entire $\deta$ coverage are also observed. While the away-side ridge mainly arises from recoiled particles from the near-side peak, the near-side ridge is a sign of long-range collective flow resulting from modulation of particle $\phi$ angular distribution. The near-side ridge is most prominent in the highest-multiplicity classes, and its strength decreases with lower event multiplicity. 
In the last panels of Figs.~\ref{fig:cf2d_pPb} and~\ref{fig:cf2d_Pbp}, correlation functions are constructed from events in the low-multiplicity class. They are used to subtract the non-flow contribution, which is described in the next section.

\subsection{Flow coefficients extraction}
\label{sec:extract_vn}

Considering a large number of collisions, the two-particle azimuthal anisotropy can be described with a Fourier transform~\cite{Voloshin1996,2012249} such that the azimuthal angle distribution of the produced pair of charged particles is written as:
\begin{equation}
   \frac{\deriv N_{\rm pair}}{\deriv \Delta\phi} = A \Big( 1 +2\sum^{\infty}_{n=1} V_{n\Delta} \cos[n (\Delta \phi) ] \Big),
   \label{eq:2pcequation}
\end{equation}
where $V_{n\Delta}$ is the Fourier coefficient of the pair distribution, and $A$ is a normalisation factor. 
Furthermore, the Fourier coefficient of a pair of particles can be written  in terms of the harmonic flow coefficients
\begin{equation}
   V_{n\Delta}(\pt^{\rm trig},\pt^{\rm assoc}) = v_n (\pt^{\rm trig}) v_n (\pt^{\rm assoc}).
   \label{eq:factorization}
\end{equation}
Therefore, the flow coefficients of particles with a specific $\pt^{\rm trig}$ are given by
\begin{equation}
    v_n (\pt^{\rm trig})
    = \frac{V_{n\Delta}(\pt^{\rm trig},\pt^{\rm assoc})}{\sqrt{V_{n\Delta}(\pt^{\rm assoc},\pt^{\rm assoc})}}.
    \label{eq:vn}
\end{equation}

\begin{figure}[!tb]
    \begin{center}
    \includegraphics[width=0.95\linewidth]{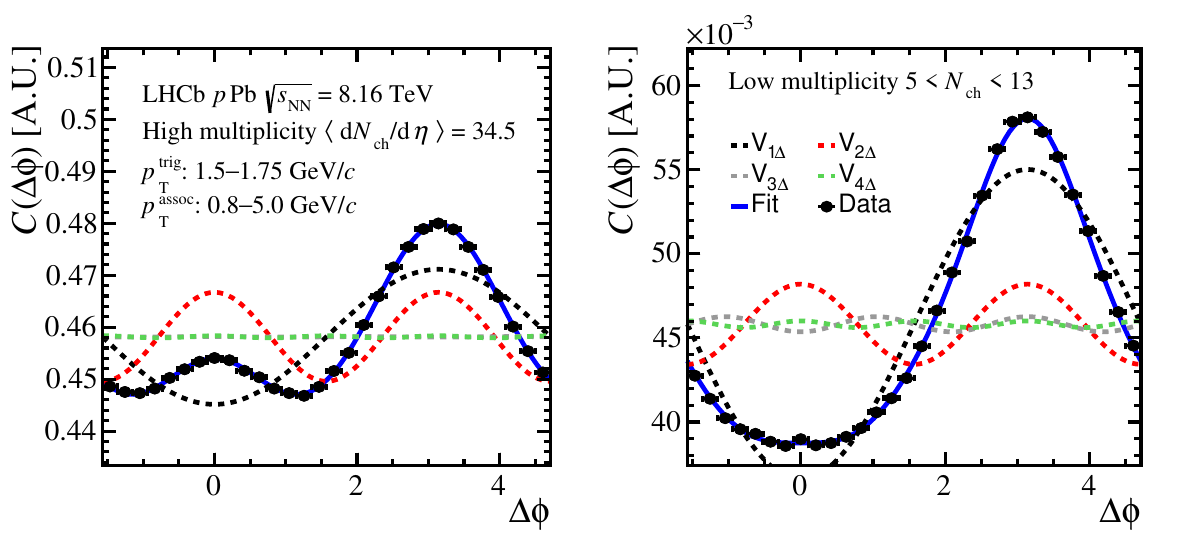}
    \vspace*{-0.5cm}
    \includegraphics[width=0.95\linewidth]{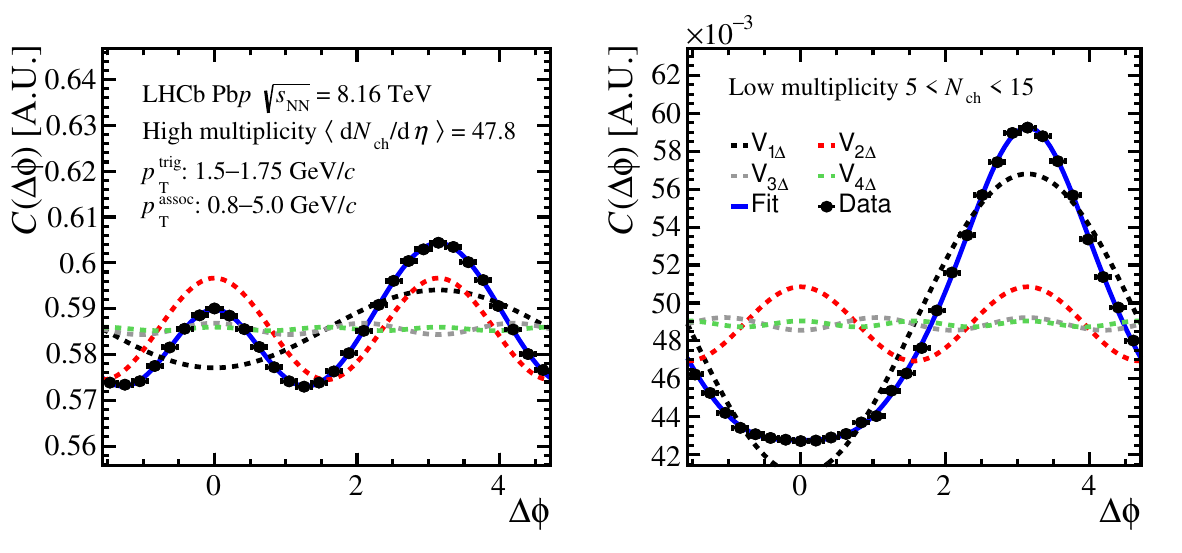}
        \vspace*{-0.5cm}
    \end{center}
    \caption{\small Examples of Fourier fits of the (top left) high- and (top right) low-multiplicity bins in $p$Pb collisions for \mbox{$1.5<p^{\text{trig}}_{\rm T}<1.75~\gevc$}, and of the (bottom left) high- and (bottom right) low-multiplicity bins in Pb$p$ collisions for \mbox{$1.5<p^{\text{trig}}_{\rm T}<1.75~\gevc$}. The different harmonics $n$ of the fits are labelled $V_{n\Delta}$.}
    \label{fig:fitsCF}
\end{figure}

The coefficients $V_{n\Delta}$ are extracted using a binned maximum-likelihood fit to the correlation functions projected onto the $\Delta \phi$ variable, using the function defined in Eq.~\ref{eq:2pcequation} up to the fourth order. 
The parameters $V_{n\Delta}$ and the normalisation $A$ are left to vary in the fit. 
Examples of Fourier fits to the correlation functions in high- and low-multiplicity in $p$Pb and Pb$p$ collisions are shown in Fig.~\ref{fig:fitsCF}.
In addition to the contributions due to collective flow effects, the two-particle correlations
also receive significant contributions from other correlation sources, such as resonances,
 minijets, and quantum correlations
~\cite{LHCb-PAPER-2023-002, PhysRevLett.59.2527,PhysRevC.89.064910,
HanburyBrown:1954amm,
Brown:1956zza,
HanburyBrown:1956bqd}. 
Such non-flow contributions are usually short-ranged. Therefore, to
limit the influence of those additional effects, the measurements aiming to study collective
flow phenomena focus on long-range correlations.

One obvious origin of the non-flow is the jet contribution that is clearly visible for \mbox{$\Delta \phi \approx 0$} and \mbox{$\Delta \eta \approx 0$}. A way to avoid this contamination is to exclude the short range (SR) region defined by \mbox{$|\Delta \eta|<1.8$}. Hence the fit is performed in the long range (LR) region, \mbox{$|\Delta \eta|>1.8$}. Additional non-flow effects, less obvious, are present at \mbox{$\Delta \phi \approx \pi$}. Therefore, methods have been developed to quantify these effects.

\begin{figure}[!tb]
    \begin{center}
    \begin{minipage}{0.48\textwidth}
        \centering
        \includegraphics[width=\linewidth]{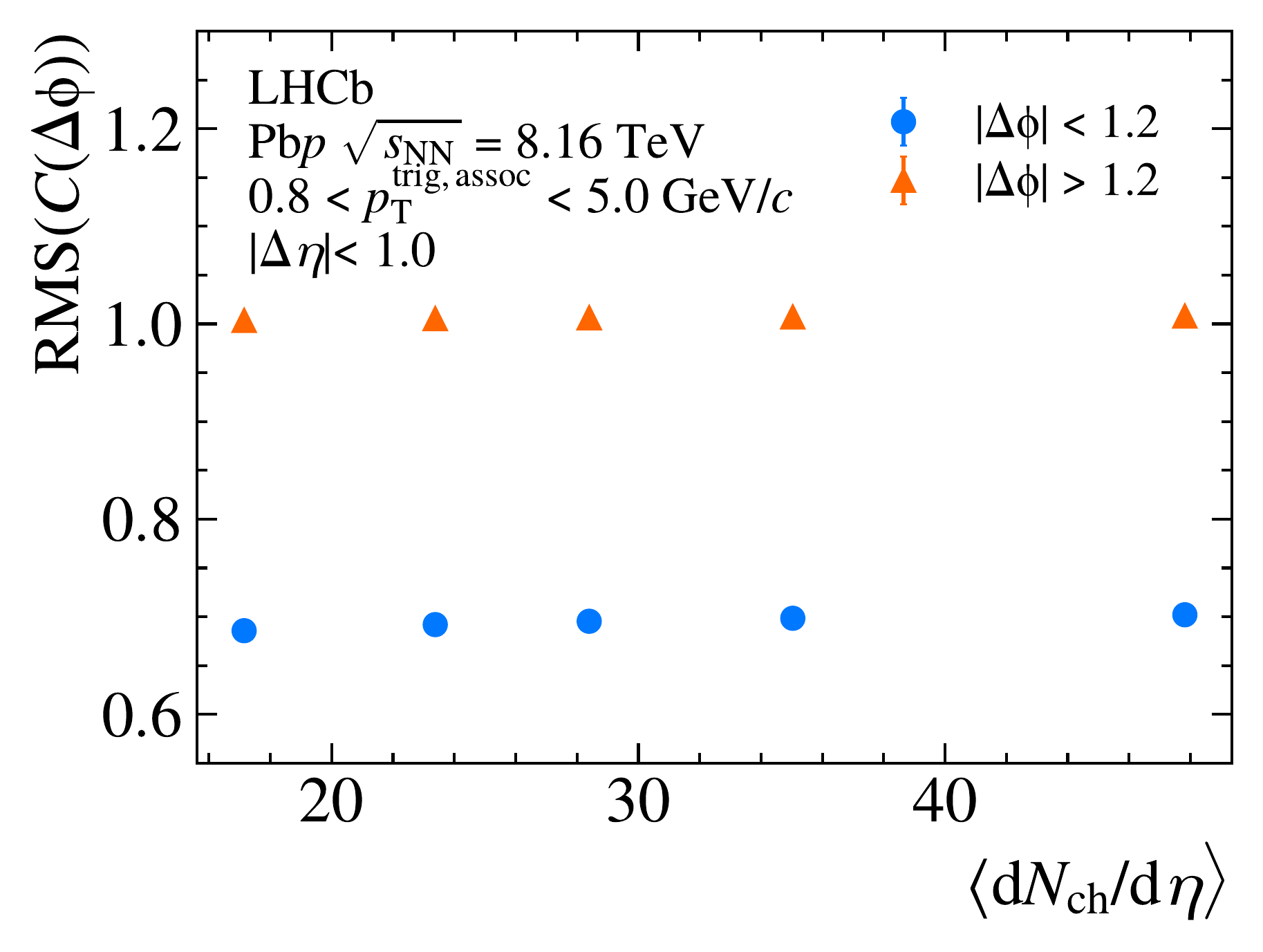} 
    \end{minipage}
    \vspace*{-0.5cm}
    \end{center}
    \caption{\small Root mean square of short-range \mbox{$C(\Delta\phi)$} in the near and away side in different Pb$p$ multiplicity bins.}
    \label{fig:SR_dphi}
\end{figure}

The method used here is taken from Ref.~\cite{CMS:2016fnw}. It is based on the assumption that the correlation coefficients ($V_{n\Delta}$) are a linear combination of flow ($c_n$) and non-flow ($d_n$) contributions such that \mbox{$V_{n\Delta} = c_n + d_n$}. Equation~\ref{eq:2pcequation} can then be rewritten as
\begin{equation}
    C(\Delta \phi) = G \Big( 1 + 2\sum^{\infty}_{n=1}(c_n + d_n)\cos[n(\Delta\phi)] \Big),
\end{equation}
 where $G$ is a normalisation factor. In this analysis the objective is to extract the $c_n$ coefficients. 
Thus $d_n$ must be subtracted from the results from the direct Fourier fit.
The strategy to estimate $d_n$ is to exploit a low-multiplicity interval where no flow is expected. Under the assumption that the shape of the non-flow correlation does not change with varying the multiplicity, $d_n$ can be written as
    \begin{equation}
       d^{\rm HM}_n = J_{\rm LM}^{\rm HM} d^{\rm LM}_n,
    \end{equation}
where LM stands for low multiplicity and HM for high multiplicity, and $J_{\rm LM}^{\rm HM}$ is the scaling factor from LM to HM. At low multiplicity flow effects are negligible, thus $d^{\rm LM}_n = V_{n\Delta}^{\rm LM}$.  
The factor $J_{\rm LM}^{\rm HM}$ can be interpreted as the relative jet yields between low and high multiplicities, expressed as
\begin{equation}
    J_{\rm LM}^{\rm HM} = \frac{N^{\rm LM} \int_{-1}^{1} (C_{SR}^{\rm HM}(\Delta\phi) - C_{LR}^{\rm HM}(\Delta\phi)) \deriv \Delta\phi}{N^{\rm HM} \int_{-1}^{1} (C_{SR}^{\rm LM}(\Delta\phi) - C_{LR}^{\rm LM}(\Delta\phi))\deriv \Delta\phi},
    \label{eq:jhmlm}
\end{equation}
where $N^{\rm LM(HM)}$ is the total integral of $C^{\rm LM(HM)}$. 
In this analysis, the HM events are taken from the five event-multiplicity classes defined in Table~\ref{tab:nch_bins}. The LM class events are selected with the criterion $5 < \nch < 13$ for the $p$Pb and $5 < \nch < 15$ for the Pb$p$ sample.
The mean charged particle density \meandndeta is $5.2 \pm 0.2$ and $5.9 \pm 0.2$ for the $p$Pb and Pb$p$ LM class, respectively.
Figure~\ref{fig:SR_dphi} displays the short range \mbox{$C(\Delta\phi)$} for \mbox{$|\deta|<1$} and the root mean square of the near and away side in the Pb$p$ configuration, showing that the jet widths are similar for the five event-multiplicity classes, and \mbox{$C(\Delta\phi)$} reaches the minimum at approximately \mbox{$\dphi = 1.2$}.
Therefore the jet region is restricted to \mbox{$\Delta \eta \in [-1.0,1.0]$} and \mbox{$\Delta \phi \in [-1.2,1.2]$}. 
The final subtracted results of elliptic and triangular flow of charged particles are named $v^{\rm sub}_2\{2\}$ and $v^{\rm sub}_3\{2\}$, respectively. The symbol $\{2\}$ indicates that the $v^{\rm sub}_{2,3}$ coefficients are measured from the two-particle correlation method, while `sub' indicates that the results are jet subtracted.

%%%%%%%%%%%%%%%%%%%%%%%%%%%%%%%%%%%%%%%%%
%%%%%%%%%%%%%%%%%%%%%%%%%%%%%%%%%%%%5

\section{Systematic uncertainties}
\label{sec:systematics}

Several sources of systematic uncertainty are considered in the final results, primarily related to the various elements of the method defined in Sec.~\ref{sec:extract_vn}. In particular, these include the effects of the low-multiplicity class definition, the long-range $|\deta|$ cut, the $|\dphi|$ range used for jet normalisation, the correction of track efficiency and contamination of fake and secondary tracks, and the uncertainty in the $\nch$ determination. A summary of the systematic uncertainties is presented in Table~\ref{tab:systtotpPb} for $p$Pb and in Table~\ref{tab:systtotPbp} for Pb$p$. The total systematic uncertainty is obtained by summing the individual components in quadrature.

To assess the impact of the low-multiplicity bin definition, alternative ranges of $\nch$ are considered. For Pb$p$ the $\nch$ ranges tested are [5,10], [10,15], [15,20], and [10,20], with the baseline range being [5,15]. For $p$Pb, the tested ranges are [5,15], [7,15], [5,10], and [10,15], with the baseline range  being [5,13]. The systematic uncertainty is quantified as the standard deviation of the results obtained across these different ranges.

The effect of the $\Delta\eta$ jet exclusion cut is tested using \mbox{$|\Delta\eta| > 1.6$} and \mbox{$|\Delta\eta| > 2.0$}. The uncertainty is calculated by taking the largest difference observed in the $v^{\text{sub}}_n\{2\}$ results relative to the baseline value of \mbox{$|\Delta\eta| > 1.8$}.

The track efficiency and contamination correction is expected to have little effect on the measurement.
However, a test is performed with and without the correction, and a systematic uncertainty is assigned as the full shift under the two alternative approaches.

The $|\dphi|$ range used to normalise the low-multiplicity jets for the subtraction method is varied, showing very little effect on the final results. Similarly, no significant effects are observed on the final results when modifying the fake track rejection criteria.

The dominant systematic uncertainty in the $\nch$ determination arises from the efficiency and purity correction applied to the tracks. A closure test using the $p$Pb simulation is performed to compare the $\nch$ value estimated with the efficiency and purity correction to the number of generated charged particles. The maximum difference of $3\%$ is assigned as an uncertainty to the mean $\nch$ values in all multiplicity classes. In the track-matching algorithm used to remove background candidates from simulated samples, occasionally a correctly reconstructed track may be mistaken for a fake track. A systematic uncertainty of 1.9\% for $p$Pb and 2.5\% for Pb$p$ collisions is assigned to account for this effect. The total systematic uncertainty of $\nch$ is 3.5\% and 3.9\% for $p$Pb and Pb$p$ collisions, respectively.

%==========================================
\begin{table}[th]
  \caption{
    \small 
    Summary of the different systematic uncertainty sources and total systematic uncertainty on the $v^{\rm sub}_{2,3}\{2\}$ determination in the $p$Pb sample 
    in bins of the mean charged-particle density $\meandndeta$ for differential and integrated $\pt^{\rm trig}$. 
    The numeric values are shown in percentage with respect to the central values.
    }
\begin{center}
\begin{tabular}{c c c c c} 
\hline
Systematic uncertainty & $\meandndeta$ & $v^{\rm sub}_2\{2\}$   & $v^{\rm sub}_2\{2\}$ & $v^{\rm sub}_3\{2\}$  \\ 
    &   & vs. \pt (\%) &   int. \pt (\%) & int. \pt (\%) \\ 
\hline
Low-multiplicity bin definition          & 34.5 & \phantom{0}1.1--6.8\phantom{00}   & 1.5 & 2.9 \\ 
                   & 25.9 & \phantom{0}1.6--13.2\phantom{0}  & 2.3 & 3.9\\ 
                   & 21.8 & \phantom{0}2.1--25.0\phantom{0}  & 3.2 & 5.2\\ 
                   & 18.2 & \phantom{0}2.7--87.9\phantom{0}  & 5.0 & 5.3\\ 
                   & 15.6 & 11.1--132.6& 9.3& 7.0\\
\hline
$\Delta\eta$ jet exclusion & 34.5 & 0.6--17.5 & 0.9 & 1.1\\ 
                   & 25.9 & 0.4--50.2 & 0.7 & 1.3\\ 
                   & 21.8 & 0.6--82.7 & 0.5 & 2.1\\ 
                   & 18.2 & 0.9--32.5 & 0.6 & 1.6\\
                   & 15.6 & 1.3--63.9 & 0.1 & 2.7\\
\hline
Track efficiency and   & 34.5 & 1.8--7.7\phantom{0}  & 3.6  & 2.4 \\ contamination correction  & 25.9 & 0.5--9.6\phantom{0}  & 5.2  & 3.2 \\ 
                   & 21.8 & 0.6--10.6 & 6.4  & 4.1 \\ 
                   & 18.2 & 0.6--16.5 & 7.5  & 3.1 \\
                   & 15.6 & 7.7--22.2 & 11.8 & 2.7 \\
\hline
Total              & 34.5 & \phantom{0}3.5--18.8\phantom{0} & 4.0   & 4.0  \\ 
                   & 25.9 & \phantom{0}4.4--51.9\phantom{0} & 5.7   & 5.2  \\ 
                   & 21.8 & \phantom{0}5.8--86.4\phantom{0} & 7.2   & 7.0  \\ 
                   & 18.2 & \phantom{0}6.5--93.7\phantom{0} & 9.0   & 6.4  \\
                   & 15.6 & \phantom{0}8.9--148.2        & 15.0  & 8.0  \\
\hline
\end{tabular}
\end{center}
\label{tab:systtotpPb}
\end{table}

%==========================================
\begin{table}[th]
  \caption{
    \small      
    Summary of the different systematic uncertainty sources and total systematic uncertainty on the $v^{\rm sub}_{2,3}\{2\}$ determination in the Pb$p$ sample 
    in bins of the mean charged-particle density $\meandndeta$ for differential and integrated $\pt^{\rm trig}$.
    The numeric values are shown in percentage with respect to the central values.
    }
\begin{center}
\begin{tabular}{c c c c c} 
\hline
Systematic uncertainty & $\meandndeta$ & $v^{\rm sub}_2\{2\}$   & $v^{\rm sub}_2\{2\}$ & $v^{\rm sub}_3\{2\}$  \\ 
      &                   & vs. \pt (\%) &   int. \pt (\%) & int. \pt (\%) \\ 

\hline
Low-multiplicity bin definition & 47.8 & 0.1--2.9\phantom{0}  &0.1 & 2.9\\
          & 35.0 & 0.3--5.3\phantom{0}  &0.2  & 4.7\\
          & 28.4 & 0.4--8.8\phantom{0}  &0.3  & 6.9\\
          & 23.4 & 0.6--17.2 &0.5 & 8.7\\
          & 17.1 &  1.3--60.2       & 1.0 & 14.7\\
\hline
$\Delta\eta$ jet exclusion & 47.8 & 0.4--2.9\phantom{0} &0.7 & 0.2\\
                      & 35.0 & 0.1--2.4\phantom{0} &0.1& 0.9\\
                      & 28.4 & 0.2--3.6\phantom{0} &0.2& 0.9\\
                      & 23.4 & 0.6--10.4&0.3& 1.4\\
                      & 17.1 & 0.7--20.6 & 0.7    & 1.5\\
\hline
Track efficiency and  & 47.8 & 2.1--5.9\phantom{0} &  2.3& 1.6 \\
contamination correction & 35.0 & 1.5--3.9\phantom{0} & 3.6 & 1.2 \\
                 & 28.4 & 1.8--4.1\phantom{0} & 4.3 & 0.2 \\
                 & 23.4 & 1.2--5.9\phantom{0} & 5.3 & 0.5 \\
                 & 17.1 & 2.1--8.8\phantom{0} & 8.1       & 1.0 \\
\hline
Total             & 47.8 & 2.4--6.0\phantom{0} & 2.4 & 3.3\\
                 & 35.0 & 2.9--5.8\phantom{0} & 3.6 & 5.0\\
                 & 28.4 & 4.2--9.7\phantom{0} & 4.3 & 7.0\\
                 & 23.4 & 5.9--20.1& 5.3 & 8.8\\
                 & 17.1 & 8.0--63.7 & 8.2       & 14.8\\
                      
\hline
\end{tabular}
\end{center}
\label{tab:systtotPbp}
\end{table}

%################################################
\section{Results}
\label{sec:Results}

\begin{figure}[tbh]
    \begin{center}
        \includegraphics[width=0.9\linewidth]{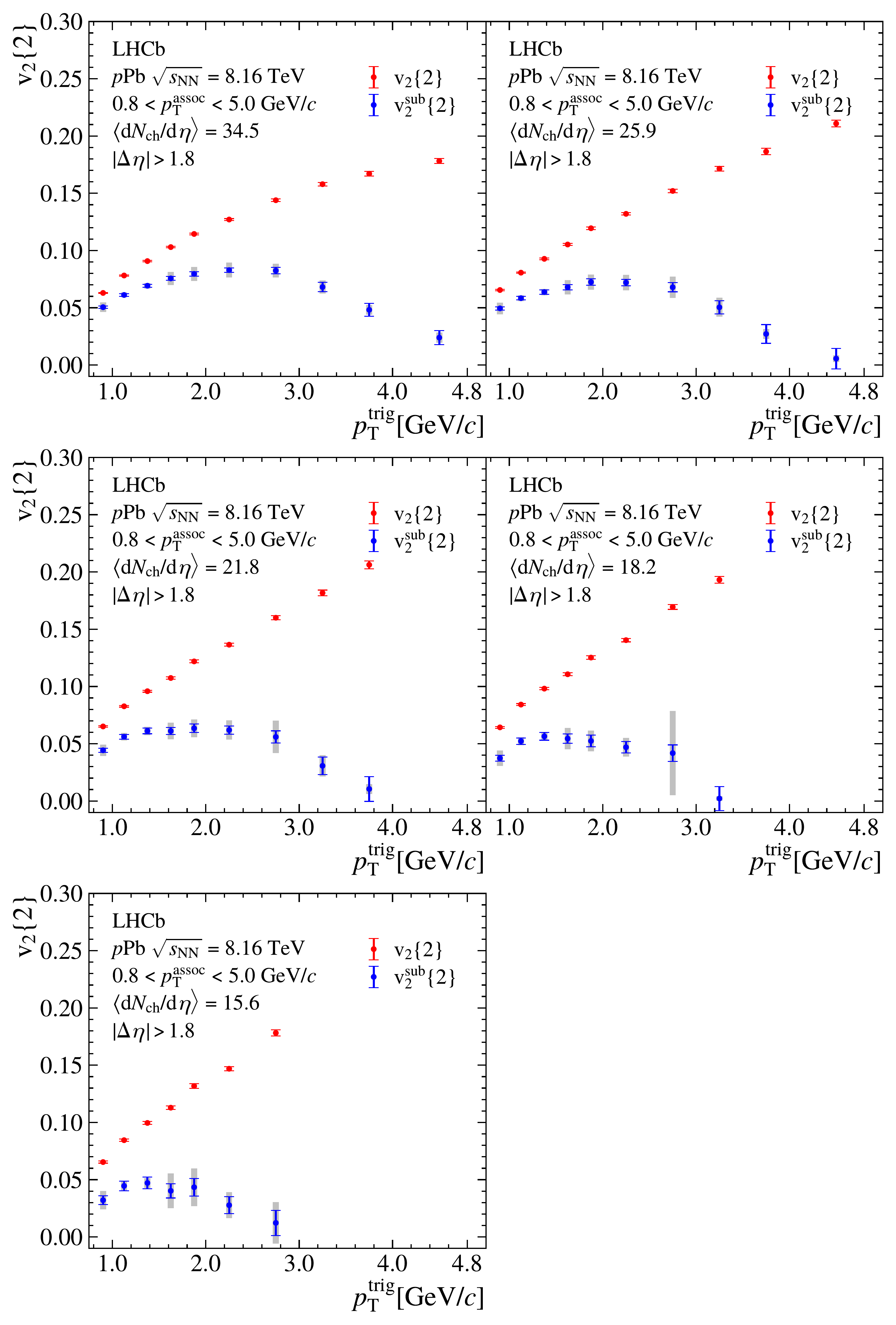}
        \vspace*{-0.5cm}
    \end{center}
    \caption{\small Results for $v_2\{2\}$ as a function of $p_{\rm T}$ for the $p$Pb sample in different $\langle \deriv N_{\mathrm{ch}}/\deriv \eta \rangle$ bins, presented before (red) and after (blue) the non-flow subtraction procedure. The vertical error bars represent statistical uncertainties while the shaded boxes represent systematic uncertainties.
    }
    \label{fig:pPbresults}
\end{figure}

\begin{figure}[tbh]
    \begin{center}
        \includegraphics[width=0.9\linewidth]{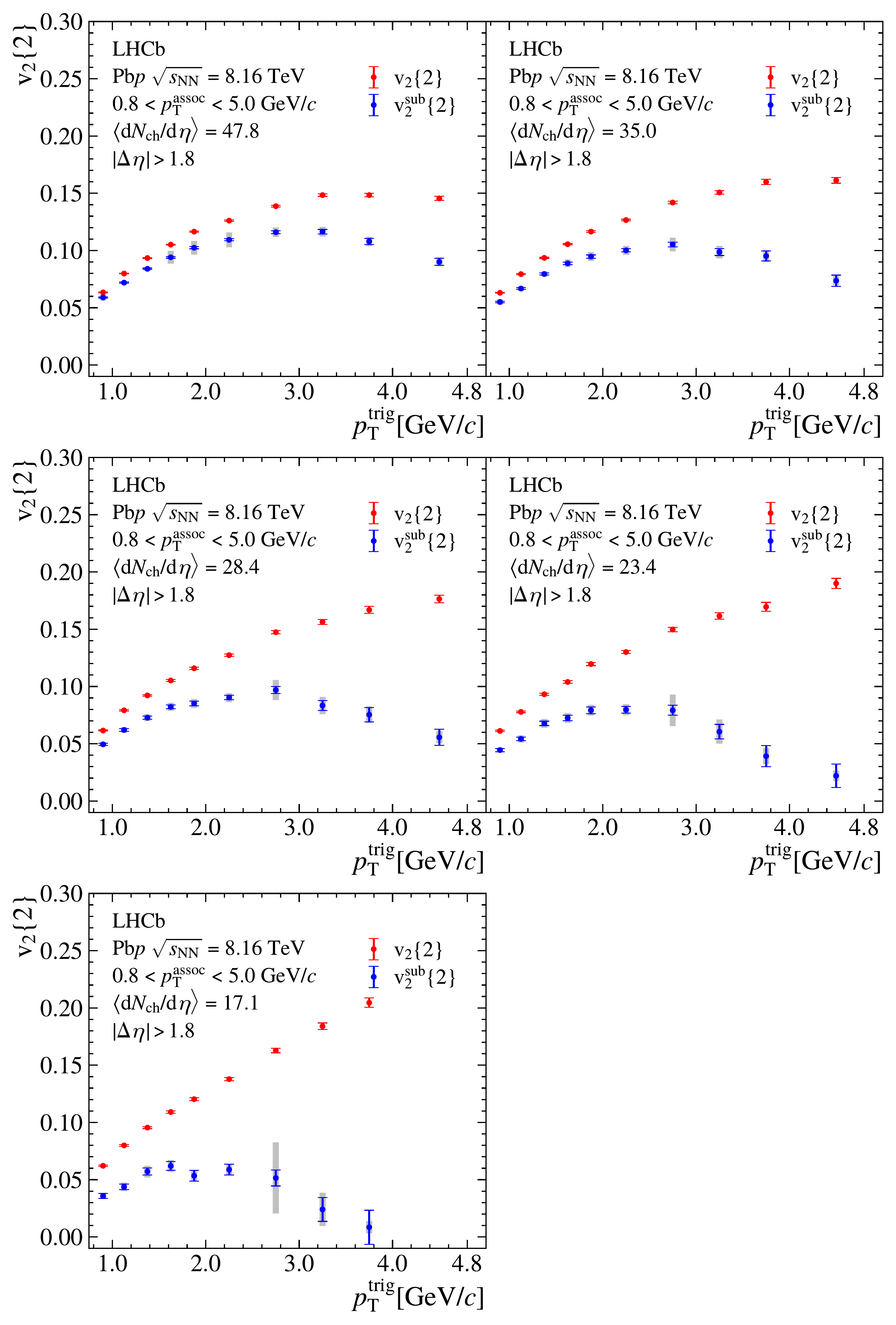}
        \vspace*{-0.5cm}
    \end{center}
    \caption{\small Results for $v_2\{2\}$ as a function of $p_{\rm T}$ for the Pb$p$ sample in different $\langle \deriv N_{\mathrm{ch}}/\deriv \eta \rangle$ bins, presented before (red) and after (blue) the non-flow subtraction procedure.
    The vertical error bars represent statistical uncertainties while the shaded boxes represent systematic uncertainties.
    }
    \label{fig:Pbpresults}
\end{figure}

The $v^{\rm sub}_2\{2\}$ results for the $p$Pb and Pb$p$ samples versus $p_{\rm T}$ in different $\meandndeta$ bins are presented in Figs.~\ref{fig:pPbresults} and~\ref{fig:Pbpresults}, respectively. The results are displayed before and after the non-flow subtraction procedure. 
For the $v_2\{2\}$ values before the subtraction, the effect arising from jets is visible as a clear increase at higher \pt. The subtraction takes care of removing the high-\pt non-flow contribution, and gives the expected trend of the elliptic flow coming from soft QCD effects, whereby $v^{\rm sub}_2\{2\}$ first increases at low \pt and starts to decrease for \pt greater than around 2.5\gevc. 
Moreover, a decrease of the flow coefficient $v^{\rm sub}_2\{2\}$ at smaller $\meandndeta$ values is clearly observed as expected. 
The detailed numerical tables of results can be found in Tables~\ref{tab:resultsPbp92}--\ref{tab:resultsv3pPbintpt} in Appendix~\ref{sec:tables_results}.

\begin{figure}[tbh]
    \begin{center}
        \includegraphics[width=0.65\linewidth]{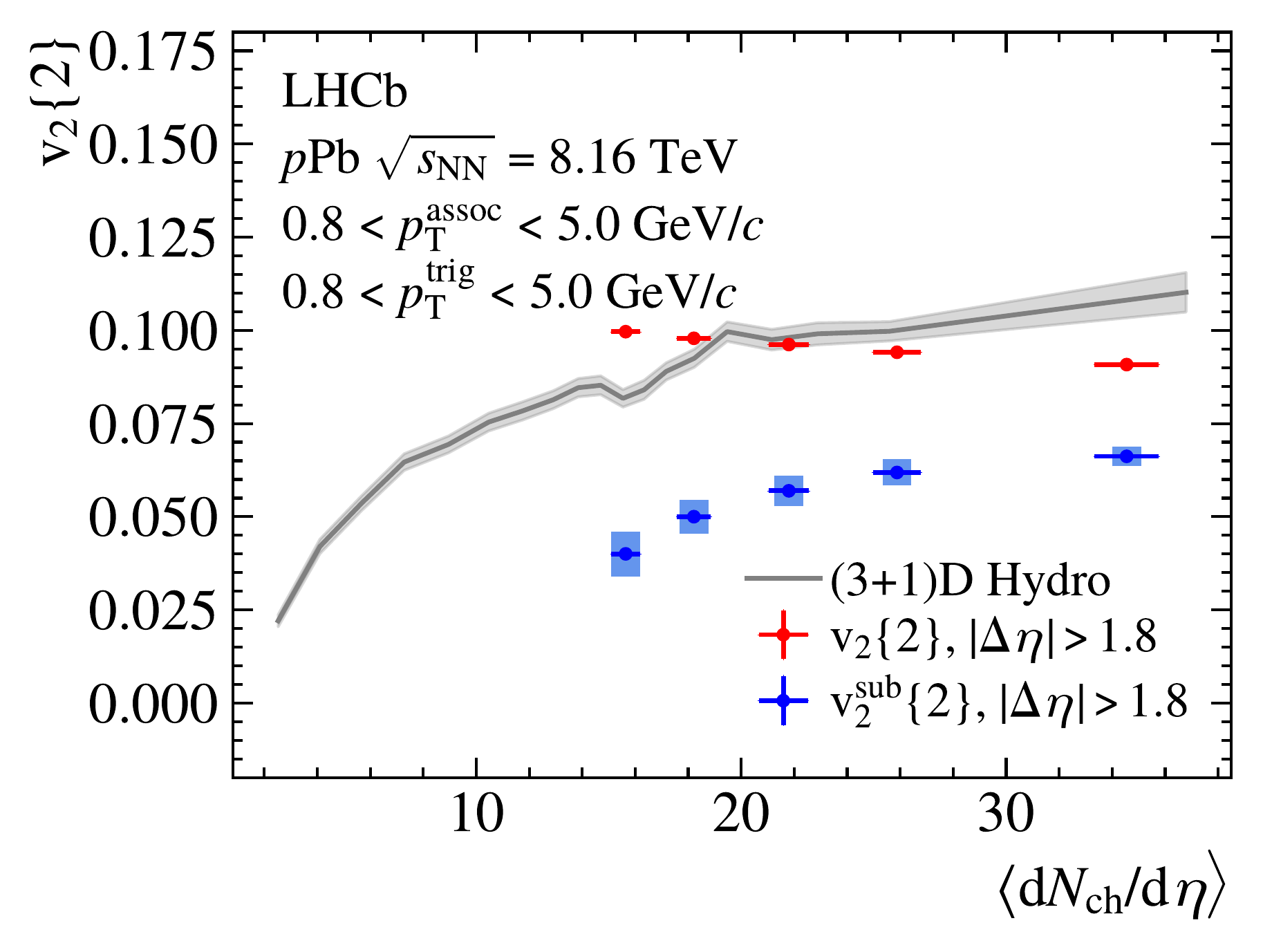}
        \vspace*{-0.5cm}
    \end{center}
    \caption{\small Results for $v_2\{2\}(|\Delta\eta|>1.8)$ as functions of $\langle \deriv N_{\mathrm{ch}}/\deriv \eta \rangle$ for the $p$Pb sample, presented before and after the non-flow subtraction procedure. 
    The vertical error bars represent statistical uncertainties, which are hidden by the data points. The shaded boxes represent systematic uncertainties.
    The results are compared with theoretical predictions based on Refs.~\cite{PhysRevLett.129.252302, PhysRevC.105.064905} for \mbox{$2.0<|\eta|<4.8$}. The shaded band represents theoretical uncertainties.
    }
    \label{fig:pPbresults_vnch}
\end{figure}

\begin{figure}[tbh]
    \begin{center}
    \includegraphics[width=0.65\linewidth]{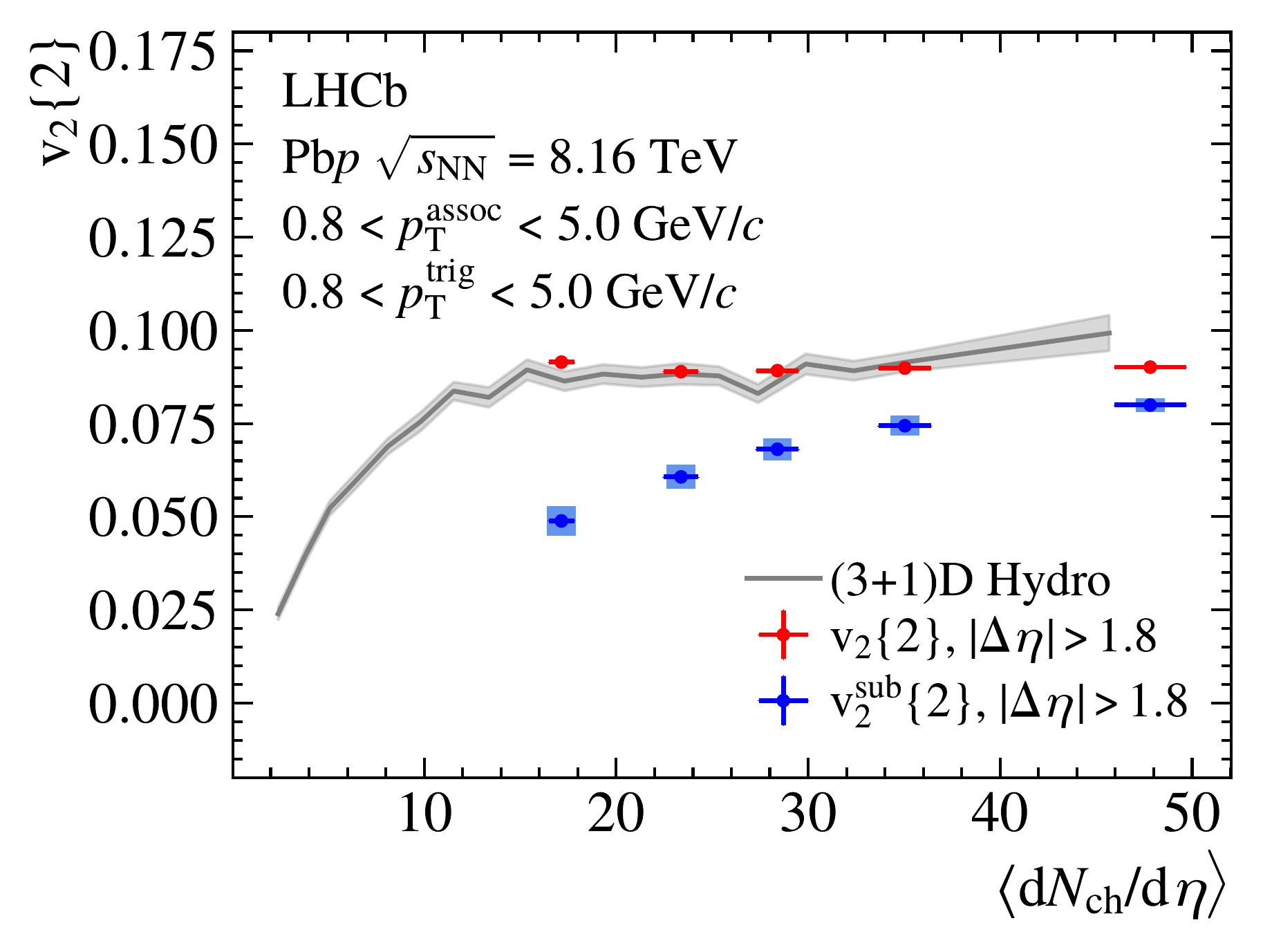}
        \vspace*{-0.5cm}
    \end{center}
    \caption{\small 
    Results for $v_2\{2\}(|\Delta\eta|>1.8
    )$ as functions of $\langle \deriv N_{\mathrm{ch}}/\deriv \eta \rangle$ for the Pb$p$ sample, presented before and after the non-flow subtraction procedure. 
    The vertical error bars represent statistical uncertainties, which are hidden by the data points. The shaded boxes represent systematic uncertainties.
    The results are compared with theoretical predictions based on Refs.~\cite{PhysRevLett.129.252302, PhysRevC.105.064905} for \mbox{$2.0<|\eta|<4.8$}. The shaded band represents theoretical uncertainties.
    }
    \label{fig:PbpresultsvNch}
\end{figure}

The results of $v_2\{2\}$ versus $\meandndeta$ for the integrated \pt range \mbox{$0.8 < \pt < 5.0 \gevc$} are presented in Figs.~\ref{fig:pPbresults_vnch} and~\ref{fig:PbpresultsvNch} for the $p$Pb and Pb$p$ data, respectively. 
There is a clear increase of $v^{\rm sub}_2\{2\}$ values with higher event multiplicity, while $v_2\{2\}$ without the subtraction reaches a plateau for the observed $\meandndeta$ range in the measurement. 
Again it is evident that the non-flow subtraction procedure has an important effect on $v^{\rm sub}_2\{2\}$.

The results 
are compared to full (3+1)D dynamical calculations coupled with a hydrodynamic and hadronic transport hybrid framework to simulate the entire dynamics of relativistic nuclear collisions~\cite{PhysRevLett.129.252302, PhysRevC.105.064905}. An important ingredient is the link made between the initial-state conditions and the later hydrodynamic evolution including energy-momentum and baryon-number conservation through Glauber-model and string-like formalism. 
The model has been calibrated using various measurements at mid-rapidity region.
The hydrodynamic parameters associated with the multiplicity distributions are determined by fitting to the multiplicity distribution data from proton-proton collisions at 7\tev, as measured by the ALICE collaboration~\cite{refId0}. 
The parameters related to the QGP viscosities are constrained by fits to the flow measurements in proton–lead collisions at $\sqsnn=5.02\tev$ from the ATLAS collaboration~\cite{PhysRevC.96.024908}, as well as to the mean transverse momentum ($\langle \pt \rangle$) in $p$Pb collisions at the LHC reported by the ALICE collaboration~\cite{201425}. 
Collective flow in the model is developed through initial geometry from Glauber model and the subsequent hydrodynamical evolution. 
The 3D model is calculated for \mbox{$2.0<|\eta|<4.8$}, 
where the predicted flow coefficients mostly agree between the $p$Pb and Pb$p$ configurations at comparable multiplicity, with larger $p$Pb values at high multiplicity.
The theoretical calculations, which must be compared with the jet subtracted results, are generally above the $v^{\rm sub}_{2}\{2\}$ data points, although the values are in agreement with the measured $v_{2}\{2\}$ before the non-flow subtraction. 
The comparison suggests that the collective flow developed in the forward rapidity region may be weaker than hydrodynamical expectations based on parametrisation tuned to data in the mid-rapidity region. 
The forward data can provide unique experimental constraints in longitudinal direction to improve the full 3D model.

\begin{figure}[tbh]
    \begin{center}
        \includegraphics[width=0.7\linewidth]{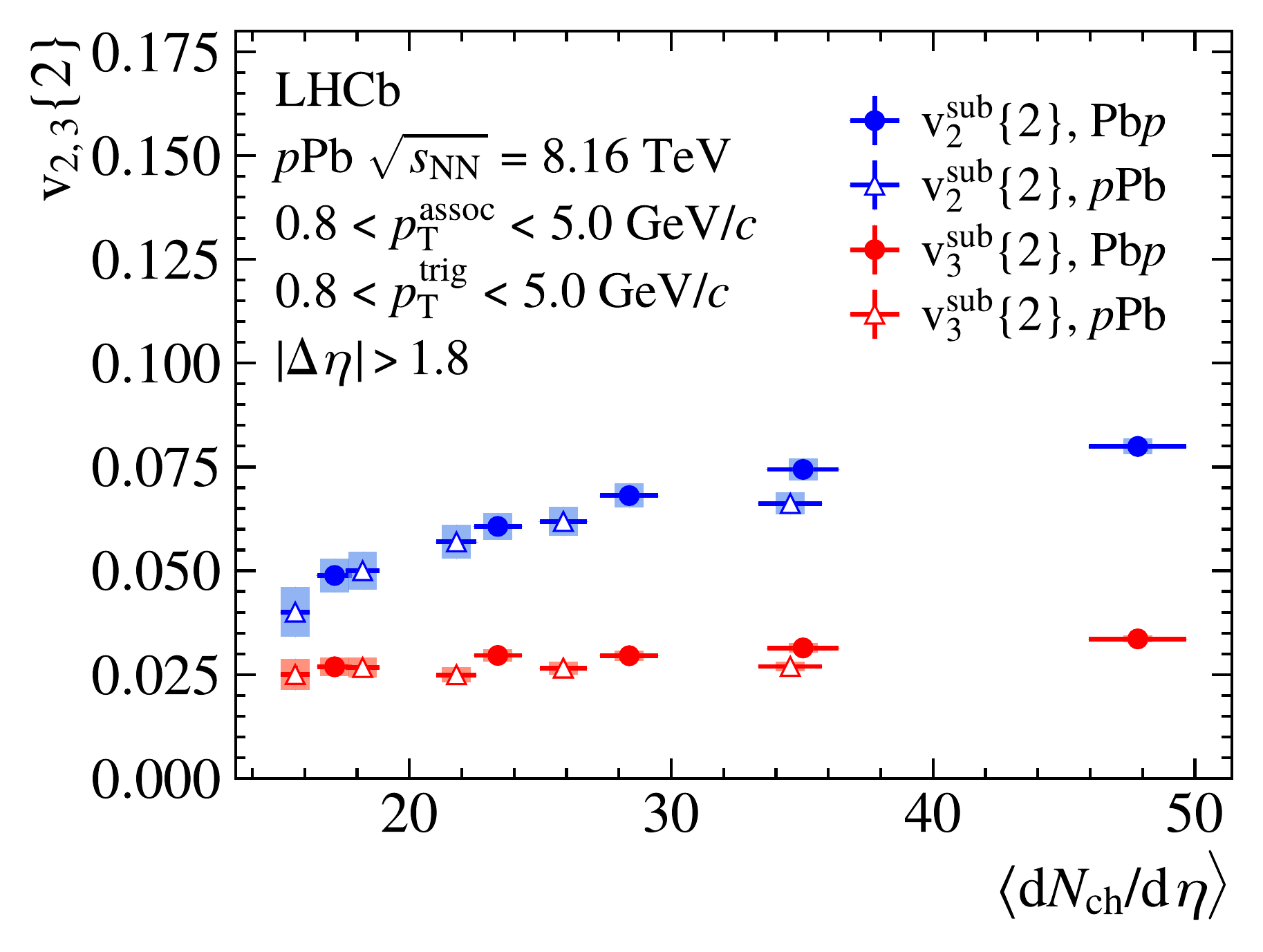}
        \vspace*{-0.5cm}
    \end{center}
    \caption{\small Results for $v^{\rm sub}_2\{2\}$ and $v^{\rm sub}_3\{2\}$ for the $p$Pb and Pb$p$ sample as functions of $\langle \deriv N_{\mathrm{ch}}/\deriv \eta \rangle$, presented after the non-flow subtraction procedure.
    The vertical error bars represent statistical uncertainties, which are hidden by the data points. The shaded boxes represent systematic uncertainties.
    }
    \label{fig:pPbPbpv2resultsnch}
\end{figure}

A direct comparison of $v_2^{\rm sub}\{2\}$ integrated in the \mbox{$0.8 < \pt < 5.0 \gevc$} range versus the event multiplicity of the collisions in both forward and backward configurations is presented in  Fig.~\ref{fig:pPbPbpv2resultsnch}. 
The forward and backward rapidity regions at \lhcb correspond to different partonic Bjorken-$x$ regions. Based on a recent \lhcb publication~\cite{LHCb-PAPER-2021-015}, in the forward region, the prompt charged hadron data probe the small Bjorken-$x$ region at approximately \mbox{$10^{-6} < x < 10^{-4}$} depending on the particle \pt, where gluon saturation may theoretically take place. In the backward region, the data cover the large Bjorken-$x$ region at approximately \mbox{$10^{-3} < x < 10^{-1}$}, where gluon saturation is not expected.  
The comparison in Fig.~\ref{fig:pPbPbpv2resultsnch} shows that the $v_2^{\rm sub}\{2\}$ values from forward and backward rapidity regions are compatible within 2.2 standard deviations at similar event multiplicity \meandndeta, 
although the particles may be affected by different initial-state effects.
This suggests that final-state effects may be dominant over initial-state effects in the origin of flow in small systems.

\begin{figure}[tbh]
    \begin{center}
        \includegraphics[width=\linewidth]{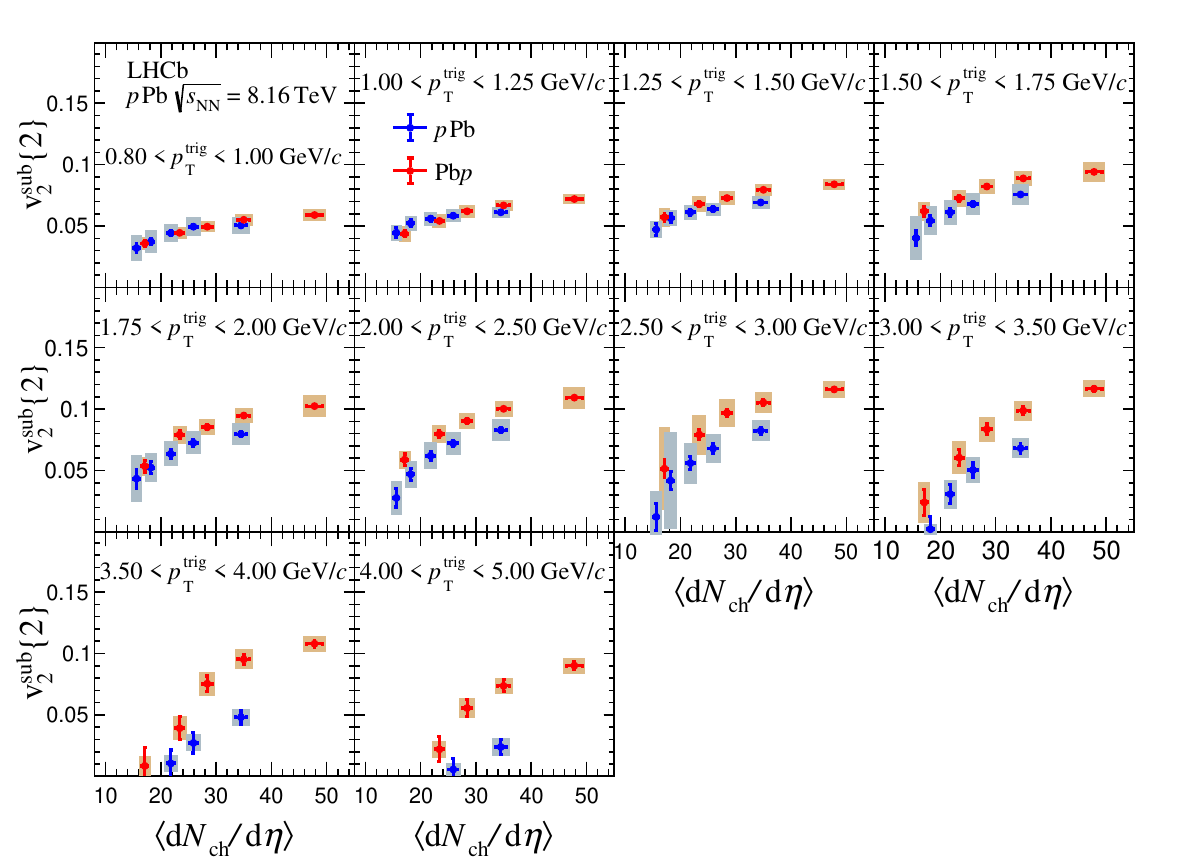}
        \vspace*{-0.5cm}
    \end{center}
    \caption{\small Results for $v^{\rm sub}_2\{2\}$(\mbox{$|\Delta\eta|>1.8$}) for the $p$Pb and Pb$p$ samples as functions of $\langle \deriv N_{\mathrm{ch}}/\deriv \eta \rangle$ in different $\pt^{\rm trig}$ ranges, presented after the non-flow subtraction procedure.
    The vertical error bars represent statistical uncertainties while the shaded boxes represent systematic uncertainties.}
    \label{fig:pPbPbpv2resultsnch_ptbins}
\end{figure}

Figure~\ref{fig:pPbPbpv2resultsnch_ptbins} compares $v_2^{\rm sub}\{2\}$ as a function of \meandndeta in forward and backward rapidities for different trigger \pt ranges. In the lower \pt bins, the $v_2^{\rm sub}\{2\}$ values in the forward and backward regions are consistent at similar \meandndeta. However, at higher \pt, the difference between $p$Pb and Pb$p$ grows significantly, reaching around five standard deviations at \mbox{$3.5 < \pt < 5.0\gevc$} and \mbox{$\meandndeta \sim 35$}. In the backward data the $v_2^{\rm sub}\{2\}$ values are systematically larger than those in the forward data. 
This could indicate some initial-state effect showing up only for high-$Q^2$ processes or the presence of some remaining non-flow effects.

\begin{figure}[tbh]
    \begin{center}
    \includegraphics[width=0.65\linewidth]{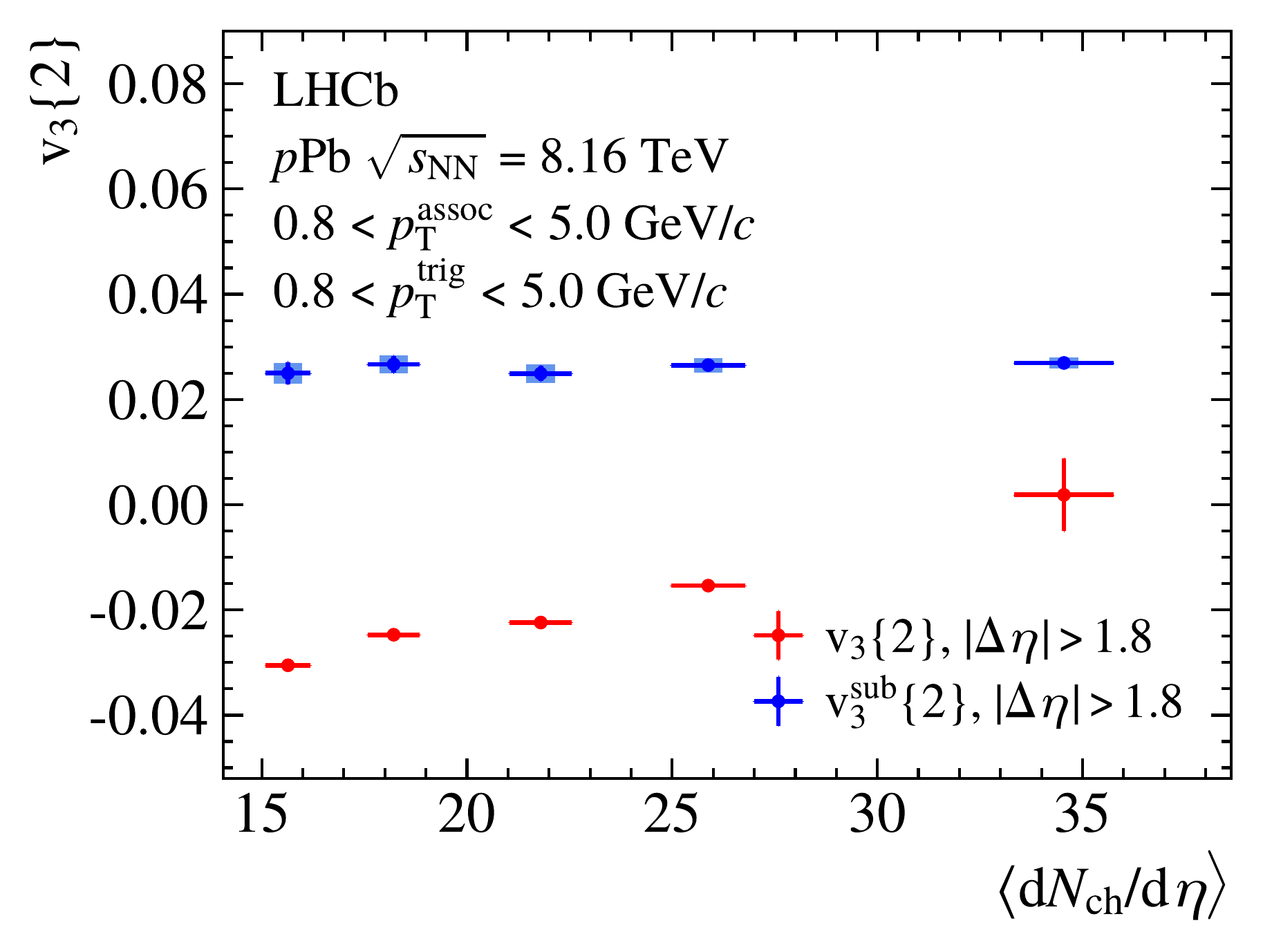}
        \vspace*{-0.5cm}
    \end{center}
    \caption{\small Results for $v_3\{2\}$(\mbox{$|\Delta\eta|>1.8$}) for the $p$Pb sample as functions of $\langle \deriv N_{\mathrm{ch}}/\deriv \eta \rangle$, presented before (red) and after (blue) the non-flow subtraction procedure. The vertical error bars represent statistical uncertainties while the shaded boxes represent systematic uncertainties.
    }
    \label{fig:pPbv3resultsvNch}
\end{figure}

\begin{figure}[tbh]
    \begin{center}
    \includegraphics[width=0.65\linewidth]{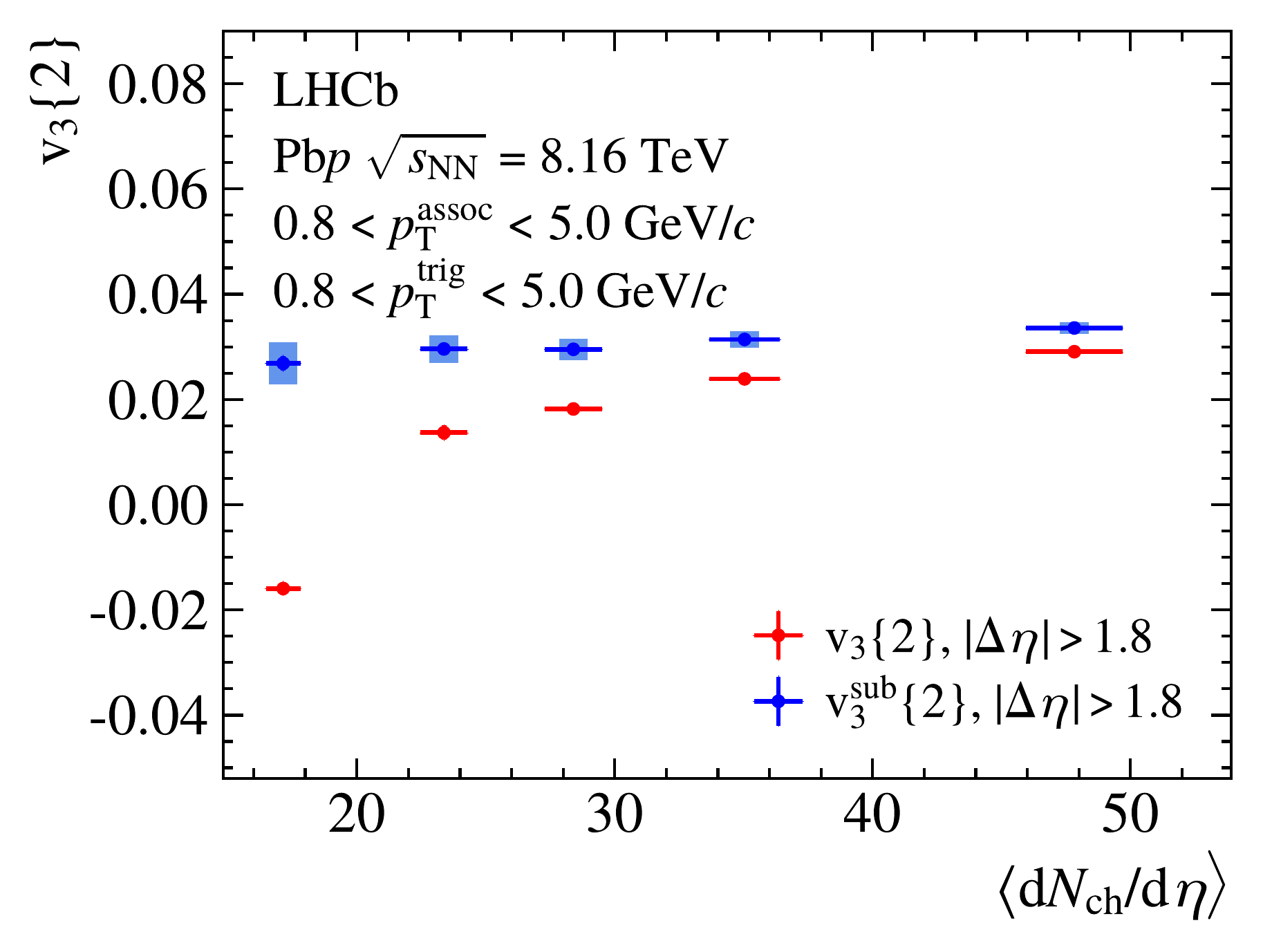}
        \vspace*{-0.5cm}
    \end{center}
    \caption{\small Results for
    $v_3\{2\}$(\mbox{$|\Delta\eta|>1.8$}) for the Pb$p$ sample as functions of $\langle \deriv N_{\mathrm{ch}}/\deriv \eta \rangle$, presented before (red) and after (blue) the non-flow subtraction procedure. The vertical error bars represent statistical uncertainties, which are hidden by the data points. The shaded boxes represent systematic uncertainties.
    }
    \label{fig:Pbpv3resultsvNch}
\end{figure}

The results corresponding to the triangular flow harmonic $v_{3}\{2\}$ and $v^{\rm sub}_{3}\{2\}$ from the $p$Pb and Pb$p$ samples for the integrated \pt range \mbox{$0.8 < \pt < 5.0\gevc$} are presented in Figs.~\ref{fig:pPbv3resultsvNch} and~\ref{fig:Pbpv3resultsvNch}, respectively. 
While the $v_{3}\{2\}$ values before the non-flow subtraction differ substantially in the forward and backward data, the subtracted $v^{\rm sub}_{3}\{2\}$ values are similar in both rapidity regions. The change of sign of $v_{3}\{2\}$ reflects the type of correlation: an anticorrelation originating from jets and a positive correlation from flow effects. As the multiplicity increases, the contribution from pure flow correlations becomes more dominant. In contrast to the $v_2^{\rm sub}\{2\}$ case, $v^{\rm sub}_{3}\{2\}$ does not exhibit strong dependence on the event multiplicity. Both $p$Pb and Pb$p$ $v^{\rm sub}_{3}\{2\}$ results are compatible with a flat trend across the observed \meandndeta ranges, while the Pb$p$ data show a hint of increase at high multiplicity with a positive slope of \mbox{$(1.9 \pm 0.5) \times 10^{-4}$}. 

%################################################
\section{Conclusion}
\label{sec:conclusion}

The elliptic and triangular flow of charged particles has been measured in $p$Pb collisions recorded by the LHCb experiment in both forward and backward configurations. The measurements are performed as a function of charged-particle multiplicity and transverse momentum for the elliptic flow. As expected, the elliptic flow increases with multiplicity. However, for a given multiplicity no significant differences are observed between the forward and backward configurations. This suggests that initial-state correlations do not play a significant role in the development of the elliptic flow. The triangular flow remains constant across multiplicities and is compatible between the two configurations. The measured elliptic flow is in general smaller than calculations using a full (3+1)D dynamical framework. 
The data can also be compared to models incorporating a 3D gluon saturation initial state framework in the future.
The result in the forward rapidity region helps to understand the origin of particle flow in small systems, and provides experimental inputs to three-dimensional theoretical models that include longitudinal dynamics.

%% file: acknowledgements.tex
\section*{Acknowledgements}
%
% These Acknowledgements valid from 3-May-2019
%
\noindent We express our gratitude to our colleagues in the CERN
accelerator departments for the excellent performance of the LHC. We
thank the technical and administrative staff at the LHCb
institutes.
We acknowledge support from CERN and from the national agencies:
ARC (Australia);
CAPES, CNPq, FAPERJ and FINEP (Brazil); 
MOST and NSFC (China); 
CNRS/IN2P3 (France); 
BMBF, DFG and MPG (Germany); 
INFN (Italy); 
NWO (Netherlands); 
MNiSW and NCN (Poland); 
MCID/IFA (Romania); 
%MSHE (Russia); 
MICIU and AEI (Spain);
SNSF and SER (Switzerland); 
NASU (Ukraine); 
STFC (United Kingdom); 
DOE NP and NSF (USA).
%%%%%%%%%%%%%%%%%%%%%%%%%%%%%%%%%%%%%%%%%%%%%
We acknowledge the computing resources that are provided by ARDC (Australia), 
CBPF (Brazil),
CERN, 
IHEP and LZU (China),
IN2P3 (France), 
KIT and DESY (Germany), 
INFN (Italy), 
SURF (Netherlands),
Polish WLCG (Poland),
IFIN-HH (Romania), 
%RRCKI and Yandex LLC (Russia), 
PIC (Spain), CSCS (Switzerland), 
and GridPP (United Kingdom).
%%%%%%%%%%%%%%%%%%%%%%%%%%%%%%%%%%%%%%%%%%
We are indebted to the communities behind the multiple open-source
software packages on which we depend.
%%%%%%%%%%%%%%%%%%%%%%%%%%%%%%%%%%%%%%%%%%
Individual groups or members have received support from
%ARC and ARDC (Australia); % moved to national 16/01
Key Research Program of Frontier Sciences of CAS, CAS PIFI, CAS CCEPP, 
Fundamental Research Funds for the Central Universities,  and Sci.\ \& Tech.\ Program of Guangzhou (China);
Minciencias (Colombia);
EPLANET, Marie Sk\l{}odowska-Curie Actions, ERC and NextGenerationEU (European Union);
A*MIDEX, ANR, IPhU and Labex P2IO, and R\'{e}gion Auvergne-Rh\^{o}ne-Alpes (France);
%RFBR, RSF and Yandex LLC (Russia);
Alexander-von-Humboldt Foundation (Germany);
ICSC (Italy); 
%GVA, XuntaGal, GENCAT, Inditex, InTalent and Prog.~Atracci\'on Talento, CM (Spain);
Severo Ochoa and Mar\'ia de Maeztu Units of Excellence, GVA, XuntaGal, GENCAT, InTalent-Inditex and Prog.~Atracci\'on Talento CM (Spain);
SRC (Sweden);
the Leverhulme Trust, the Royal Society and UKRI (United Kingdom).

%% file: appendix.tex
% ===============================================================================
% Purpose: appendix to the standard template: standard symbol alises from Ulrik
% Author: Tomasz Skwarnicki
% Created on: 2009-09-24
% ===============================================================================

\clearpage
\section*{Appendix}

\appendix

\section{Result tables}
\label{sec:tables_results}

Numeric values of $v^{\rm sub}_2\{2\}$ and $v^{\rm sub}_3\{2\}$ in \pt and $\meandndeta$ bins for $p$Pb and Pb$p$ samples are presented in Tables~\ref{tab:resultsPbp92} to \ref{tab:resultsv3pPbintpt}.

\begin{table}[h]
\caption{
\small 
    Results of $v^{\rm sub}_2\{2\}$ from the Pb$p$ samples for the bin \mbox{$\meandndeta=47.8$.}}
    \centering
    \sisetup{table-align-uncertainty=true, separate-uncertainty=true}
    \begin{tabular}{l 
                    S[table-format=1.4]
                    @{\,\( \pm \)\,} S[table-format=1.4] 
                    @{\,\( \pm \)\,} S[table-format=1.4]}
        \toprule
        \multicolumn{1}{c}{$p^{\rm trig}_{\rm T} [\gevc]$} & 
        {$v^{\rm sub}_{2}\{2\}$} & 
        {stat} & 
        {syst} \\
        \midrule
        0.80--1.00  & 0.0590 & 0.0004 & 0.0027 \\
        1.00--1.25 & 0.0720 & 0.0005 & 0.0017 \\
        1.25--1.50  & 0.0840 & 0.0006 & 0.0020 \\
        1.50--1.75 & 0.0941 & 0.0008 & 0.0056 \\
        1.75--2.00  & 0.1024 & 0.0009 & 0.0061 \\
        2.00--2.50  & 0.1093 & 0.0009 & 0.0065 \\
        2.50--3.00  & 0.1160 & 0.0013 & 0.0043 \\
        3.00--3.50  & 0.1165 & 0.0019 & 0.0042 \\
        3.50--4.00  & 0.1079 & 0.0027 & 0.0041 \\
        4.00--5.00  & 0.0901 & 0.0031 & 0.0039 \\
        \bottomrule
    \end{tabular}
    \label{tab:resultsPbp92}
\end{table}

\begin{table}[h]
  \caption{
    \small 
    Results of $v^{\rm sub}_2\{2\}$ from the Pb$p$ samples for the bin \mbox{$\meandndeta=35.0$}.}
    \centering
    \sisetup{table-align-uncertainty=true, separate-uncertainty=true}
    \begin{tabular}{l 
                    S[table-format=1.4]
                    @{\,\( \pm \)\,} S[table-format=1.4] 
                    @{\,\( \pm \)\,} S[table-format=1.4]}
        \toprule
        \multicolumn{1}{c}{$p^{\rm trig}_{\rm T} [\gevc]$} & 
        {$v^{\rm sub}_{2}\{2\}$} & 
        {stat} & 
        {syst} \\
        \midrule
        0.80--1.00  & 0.0551 & 0.0007 & 0.0025 \\
        1.00--1.25 & 0.0668 & 0.0008 & 0.0020 \\
        1.25--1.50  & 0.0795 & 0.0009 & 0.0023 \\
        1.50--1.75 & 0.0888 & 0.0011 & 0.0035 \\
        1.75--2.00  & 0.0947 & 0.0014 & 0.0039 \\
        2.00--2.50  & 0.1002 & 0.0014 & 0.0040 \\
        2.50--3.00  & 0.1053 & 0.0021 & 0.0060 \\
        3.00--3.50  & 0.0987 & 0.0030 & 0.0055 \\
        3.50--4.00  & 0.0952 & 0.0043 & 0.0055 \\
        4.00--5.00  & 0.0736 & 0.0049 & 0.0042 \\
        \bottomrule
    \end{tabular}
    \label{tab:resultsPbp35}
\end{table}

\begin{table}[h]
  \caption{
    \small 
    Results of $v^{\rm sub}_2\{2\}$ from the Pb$p$ samples for the bin \mbox{$\meandndeta=28.4$}.}
    \centering
    \sisetup{table-align-uncertainty=true, separate-uncertainty=true}
    \begin{tabular}{l 
                    S[table-format=1.4] 
                    @{\,\( \pm \)\,} S[table-format=1.4] 
                    @{\,\( \pm \)\,} S[table-format=1.4]}
        \toprule
        \multicolumn{1}{c}{$p^{\rm trig}_{\rm T} [\gevc]$} & 
        {$v^{\rm sub}_{2}\{2\}$} & 
        {stat} & 
        {syst} \\
        \midrule
        0.80--1.00  & 0.0495 & 0.0010 & 0.0021 \\
        1.00--1.25 & 0.0620 & 0.0011 & 0.0026 \\
        1.25--1.50  & 0.0728 & 0.0013 & 0.0031 \\
        1.50--1.75 & 0.0822 & 0.0016 & 0.0036 \\
        1.75--2.00  & 0.0852 & 0.0020 & 0.0039 \\
        2.00--2.50  & 0.0904 & 0.0020 & 0.0040  \\
        2.50--3.00  & 0.0969 & 0.0030 & 0.0088 \\
        3.00--3.50  & 0.0834 & 0.0044 & 0.0076 \\
        3.50--4.00  & 0.0754 & 0.0063 & 0.0073 \\
        4.00--5.00  & 0.0556 & 0.0070 & 0.0054 \\
        \bottomrule
    \end{tabular}
    \label{tab:resultsPbp28}
\end{table}

\begin{table}[h]
  \caption{
    \small 
    Results of $v^{\rm sub}_2\{2\}$ from the Pb$p$ samples for the bin \mbox{$\meandndeta=23.4$}.}
    \centering
    \sisetup{table-align-uncertainty=true, separate-uncertainty=true}
    \begin{tabular}{l 
                    S[table-format=1.4] 
                    @{\,\( \pm \)\,} S[table-format=1.4] 
                    @{\,\( \pm \)\,} S[table-format=1.4]}
        \toprule
        \multicolumn{1}{c}{$p^{\rm trig}_{\rm T} [\gevc]$} & 
        {$v^{\rm sub}_{2}\{2\}$} & 
        {stat} & 
        {syst} \\
        \midrule
        0.80--1.00  & 0.0445 & 0.0014 & 0.0022 \\
        1.00--1.25 & 0.0542 & 0.0015 & 0.0032 \\
        1.25--1.50  & 0.0678 & 0.0019 & 0.0040 \\
        1.50--1.75 & 0.0725 & 0.0024 & 0.0043 \\
        1.75--2.00  & 0.0792 & 0.0029 & 0.0048 \\
        2.00--2.50  & 0.0797 & 0.0029 & 0.0048 \\
        2.50--3.00  & 0.0792 & 0.0043 & 0.0138 \\
        3.00--3.50  & 0.0605 & 0.0064 & 0.0106 \\
        3.50--4.00  & 0.0392 & 0.0092 & 0.0071 \\
        4.00--5.00  & 0.0220 & 0.0102 & 0.0044 \\
        \bottomrule
    \end{tabular}
    \label{tab:resultsPbp23}
\end{table}

\begin{table}[h]
  \caption{
    \small 
    Results of $v^{\rm sub}_2\{2\}$ from the Pb$p$ samples for the bin \mbox{$\meandndeta=17.1$}.}
    \centering
    \sisetup{table-align-uncertainty=true, separate-uncertainty=true}
    \begin{tabular}{l 
                    S[table-format=1.4] 
                    @{\,\( \pm \)\,} S[table-format=1.4] 
                    @{\,\( \pm \)\,} S[table-format=1.4]}
        \toprule
        \multicolumn{1}{c}{$p^{\rm trig}_{\rm T} [\gevc]$} & 
        {$v^{\rm sub}_{2}\{2\}$} & 
        {stat} & 
        {syst} \\
        \midrule
        0.80--1.00  & 0.0358 & 0.0022 & 0.0029 \\
        1.00--1.25 & 0.0438 & 0.0024 & 0.0039 \\
        1.25--1.50  & 0.0571 & 0.0030 & 0.0051 \\
        1.50--1.75 & 0.0620 & 0.0038 & 0.0052 \\
        1.75--2.00  & 0.0535 & 0.0047 & 0.0052 \\
        2.00--2.50  & 0.0588 & 0.0046 & 0.0048 \\
        2.50--3.00  & 0.0515 & 0.0070 & 0.0311 \\
        3.00--3.50  & 0.0240 & 0.0104 & 0.0145 \\
        3.50--4.00  & 0.0085 & 0.0149 & 0.0054 \\
        \bottomrule
    \end{tabular}
    \label{tab:resultsPbp17}
\end{table}

\begin{table}[h]
  \caption{
    \small 
    Results of $v^{\rm sub}_2\{2\}$ from the $p$Pb samples \mbox{$\meandndeta=34.5$}. }
    \centering
    \sisetup{table-align-uncertainty=true, separate-uncertainty=true}
    \begin{tabular}{l 
                    S[table-format=1.4] 
                    @{\,\( \pm \)\,} S[table-format=1.4] 
                    @{\,\( \pm \)\,} S[table-format=1.4]}
        \toprule
        \multicolumn{1}{c}{$p^{\rm trig}_{\rm T} [\gevc]$} & 
        {$v^{\rm sub}_{2}\{2\}$} & 
        {stat} & 
        {syst} \\
        \midrule
        0.80--1.00  & 0.0505 & 0.0010 & 0.0042 \\
        1.00--1.25 & 0.0612 & 0.0011 & 0.0021 \\
        1.25--1.50  & 0.0692 & 0.0013 & 0.0024 \\
        1.50--1.75 & 0.0757 & 0.0016 & 0.0058 \\
        1.75--2.00  & 0.0796 & 0.0020 & 0.0061 \\
        2.00--2.50  & 0.0829 & 0.0019 & 0.0065 \\
        2.50--3.00  & 0.0824 & 0.0028 & 0.0059 \\
        3.00--3.50  & 0.0681 & 0.0040 & 0.0057 \\
        3.50--4.00  & 0.0482 & 0.0057 & 0.0043 \\
        4.00--5.00  & 0.0239 & 0.0062 & 0.0045 \\
        \bottomrule
    \end{tabular}
    \label{tab:resultspPb34}
\end{table}

\begin{table}[h]
  \caption{
    \small 
    Results of $v^{\rm sub}_2\{2\}$ from the $p$Pb samples for the bin \mbox{$\meandndeta=25.9$}. }
    \centering
    \sisetup{table-align-uncertainty=true, separate-uncertainty=true}
    \begin{tabular}{l 
                    S[table-format=1.4] 
                    @{\,\( \pm \)\,} S[table-format=1.4] 
                    @{\,\( \pm \)\,} S[table-format=1.4]}
        \toprule
        \multicolumn{1}{c}{$p^{\rm trig}_{\rm T} [\gevc]$} & 
        {$v^{\rm sub}_{2}\{2\}$} & 
        {stat} & 
        {syst} \\
        \midrule
        0.80--1.00  & 0.0495 & 0.0014 & 0.0051 \\
        1.00--1.25 & 0.0584 & 0.0015 & 0.0026 \\
        1.25--1.50  & 0.0638 & 0.0019 & 0.0028 \\
        1.50--1.75 & 0.0680 & 0.0023 & 0.0062 \\
        1.75--2.00  & 0.0725 & 0.0028 & 0.0067 \\
        2.00--2.50  & 0.0720 & 0.0027 & 0.0068 \\
        2.50--3.00  & 0.0679 & 0.0040 & 0.0093 \\
        3.00--3.50  & 0.0504 & 0.0058 & 0.0083 \\
        3.50--4.00  & 0.0272 & 0.0082 & 0.0044 \\
        4.00--5.00  & 0.0056 & 0.0090 & 0.0029 \\
        \bottomrule
    \end{tabular}
    \label{tab:resultspPb25}
\end{table}

\begin{table}[h]
  \caption{
    \small 
    Results of $v^{\rm sub}_2\{2\}$ from the $p$Pb samples for the bin \mbox{$\meandndeta=21.8$}. }
    \centering
    \sisetup{table-align-uncertainty=true, separate-uncertainty=true}
    \begin{tabular}{l 
                    S[table-format=1.4]
                    @{\,\( \pm \)\,} S[table-format=1.4] 
                    @{\,\( \pm \)\,} S[table-format=1.4]}
        \toprule
        \multicolumn{1}{c}{$p^{\rm trig}_{\rm T} [\gevc]$} & 
        {$v^{\rm sub}_{2}\{2\}$} & 
        {stat} & 
        {syst} \\
        \midrule
        0.80--1.00  & 0.0443 & 0.0018 & 0.0051 \\
        1.00--1.25 & 0.0561 & 0.0020 & 0.0032 \\
        1.25--1.50  & 0.0612 & 0.0025 & 0.0036 \\
        1.50--1.75 & 0.0611 & 0.0030 & 0.0073 \\
        1.75--2.00  & 0.0635 & 0.0037 & 0.0077 \\
        2.00--2.50  & 0.0620 & 0.0036 & 0.0085 \\
        2.50--3.00  & 0.0560 & 0.0053 & 0.0142 \\
        3.00--3.50  & 0.0307 & 0.0076 & 0.0093 \\
        3.50--4.00  & 0.0105 & 0.0109 & 0.0044 \\
        \bottomrule
    \end{tabular}
    \label{tab:resultspPb21}
\end{table}

\begin{table}[h]
  \caption{
    \small 
    Results of $v^{\rm sub}_2\{2\}$ from the $p$Pb samples for the bin \mbox{$\meandndeta=18.2$}. }
    \centering
    \sisetup{table-align-uncertainty=true, separate-uncertainty=true}
    \begin{tabular}{l 
                    S[table-format=1.4] 
                    @{\,\( \pm \)\,} S[table-format=1.4] 
                    @{\,\( \pm \)\,} S[table-format=1.4]}
        \toprule
        \multicolumn{1}{c}{$p^{\rm trig}_{\rm T} [\gevc]$} & 
        {$v^{\rm sub}_{2}\{2\}$} & 
        {stat} & 
        {syst} \\
        \midrule
        0.80--1.00  & 0.0374 & 0.0025 & 0.0067 \\
        1.00--1.25 & 0.0522 & 0.0028 & 0.0034 \\
        1.25--1.50  & 0.0565 & 0.0034 & 0.0037 \\
        1.50--1.75 & 0.0545 & 0.0041 & 0.0094 \\
        1.75--2.00  & 0.0524 & 0.0050 & 0.0090 \\
        2.00--2.50  & 0.0469 & 0.0049 & 0.0083 \\
        2.50--3.00  & 0.0418 & 0.0072 & 0.0368 \\
        3.00--3.50  & 0.0022 & 0.0105 & 0.0020 \\
        \bottomrule
    \end{tabular}
    \label{tab:resultspPb18}
\end{table}

\begin{table}[h]
  \caption{
    \small 
    Results of $v^{\rm sub}_2\{2\}$ from the $p$Pb samples for the bin \mbox{$\meandndeta=15.6$}. }
    \centering
    \sisetup{table-align-uncertainty=true, separate-uncertainty=true}
    \begin{tabular}{l 
                    S[table-format=1.4] 
                    @{\,\( \pm \)\,} S[table-format=1.4] 
                    @{\,\( \pm \)\,} S[table-format=1.4]}
        \toprule
        \multicolumn{1}{c}{$p^{\rm trig}_{\rm T} [\gevc]$} & 
        {$v^{\rm sub}_{2}\{2\}$} & 
        {stat} & 
        {syst} \\ 
        \midrule
        0.80--1.00  & 0.0321 & 0.0038 & 0.0081 \\
        1.00--1.25 & 0.0446 & 0.0042 & 0.0040 \\
        1.25--1.50  & 0.0472 & 0.0051 & 0.0044 \\
        1.50--1.75 & 0.0403 & 0.0062 & 0.0152 \\
        1.75--2.00  & 0.0434 & 0.0077 & 0.0164 \\
        2.00--2.50  & 0.0277 & 0.0074 & 0.0114 \\
        2.50--3.00  & 0.0123 & 0.0110 & 0.0182 \\
        \bottomrule
    \end{tabular}
    \label{tab:resultspPb15}
\end{table}

\begin{table}[t]
  \caption{
    \small 
    Results of $v^{\rm sub}_2\{2\}$ from the Pb$p$ samples for different bins of $\meandndeta$. }
    \centering
    \sisetup{table-align-uncertainty=true, separate-uncertainty=true}
    \begin{tabular}{
        S[table-format=2.1] 
        S[table-format=1.4] 
        @{\,\( \pm \)\,} S[table-format=1.4] 
        @{\,\( \pm \)\,} S[table-format=1.4]}
    \toprule
    {$\meandndeta$} & 
    {$v^{\rm sub}_{2}\{2\}$} & 
    {stat} & 
    {syst} \\ 
    \midrule
    47.8 & 0.0799 & 0.0002 & 0.0019 \\
    35.0 & 0.0744 & 0.0002 & 0.0027 \\
    28.4 & 0.0681 & 0.0003 & 0.0029 \\
    23.4 & 0.0607 & 0.0005 & 0.0032 \\
    17.1 & 0.0489 & 0.0008 & 0.0040 \\
    \bottomrule
    \end{tabular}
\label{tab:resultsv2Pbpintpt}
\end{table}

\begin{table}[t]
  \caption{
    \small 
    Results of $v^{\rm sub}_3\{2\}$ from the Pb$p$ samples for different bins of $\meandndeta$. }
    \centering
    \sisetup{table-align-uncertainty=true, separate-uncertainty=true}
    \begin{tabular}{
        S[table-format=2.1] 
        S[table-format=1.4] 
        @{\,\( \pm \)\,} S[table-format=1.4] 
        @{\,\( \pm \)\,} S[table-format=1.4]}
    \toprule
    {$\meandndeta$} & 
    {$v^{\rm sub}_{3}\{2\}$} & 
    {stat} & 
    {syst} \\ 
    \midrule
    47.8 & 0.0336 & 0.0004 & 0.0011 \\
    35.0 & 0.0314 & 0.0006 & 0.0016 \\
    28.4 & 0.0295 & 0.0008 & 0.0021 \\
    23.4 & 0.0296 & 0.0010 & 0.0026 \\
    17.1 & 0.0269 & 0.0014 & 0.0040 \\
    \bottomrule
    \end{tabular}
\label{tab:resultsv3Pbpintpt}
\end{table}

\begin{table}[t]
  \caption{
    \small 
    Results of $v^{\rm sub}_2\{2\}$ from the $p$Pb samples for different bins of $\meandndeta$. }
    \centering
    \sisetup{table-align-uncertainty=true, separate-uncertainty=true}
    \begin{tabular}{
        S[table-format=2.1] 
        S[table-format=1.4] 
        @{\,\( \pm \)\,} S[table-format=1.4] 
        @{\,\( \pm \)\,} S[table-format=1.4]}
    \toprule
    {$\meandndeta$} & 
    {$v^{\rm sub}_{2}\{2\}$} & 
    {stat} & 
    {syst} \\ 
    \midrule
    34.5 & 0.0662 & 0.0003 & 0.0027 \\
    25.9 & 0.0619 & 0.0005 & 0.0035 \\
    21.8 & 0.0569 & 0.0006 & 0.0041 \\
    18.2 & 0.0500 & 0.0009 & 0.0045 \\
    15.6 & 0.0400 & 0.0013 & 0.0060 \\
    \bottomrule
    \end{tabular}
\label{tab:resultsv2pPbintpt}
\end{table}

\begin{table}[t]
  \caption{
    \small 
    Results of $v^{\rm sub}_3\{2\}$ from the $p$Pb samples for different bins of $\meandndeta$. }
    \centering
    \sisetup{table-align-uncertainty=true, separate-uncertainty=true}
    \begin{tabular}{
        S[table-format=2.1] 
        S[table-format=1.4] 
        @{\,\( \pm \)\,} S[table-format=1.4] 
        @{\,\( \pm \)\,} S[table-format=1.4]}
    \toprule
    {$\meandndeta$} & 
    {$v^{\rm sub}_{3}\{2\}$} & 
    {stat} & 
    {syst} \\ 
    \midrule
    34.5 & 0.0269 & 0.0008 & 0.0011 \\
    25.9 & 0.0265 & 0.0011 & 0.0014 \\
    21.8 & 0.0249 & 0.0015 & 0.0017 \\
    18.2 & 0.0267 & 0.0016 & 0.0017 \\
    15.6 & 0.0250 & 0.0021 & 0.0020 \\
    \bottomrule
    \end{tabular}
\label{tab:resultsv3pPbintpt}
\end{table}

%% file: main.bbl
\ifx\mcitethebibliography\mciteundefinedmacro
\PackageError{LHCb.bst}{mciteplus.sty has not been loaded}
{This bibstyle requires the use of the mciteplus package.}\fi
\providecommand{\href}[2]{#2}

%% file: Authorship_LHCb-PAPER-2025-003.tex
% LHCb collaboration author list
% Data extracted on February 11th, 2025 at 4:02pm for paper reference LHCb-PAPER-2025-002
\centerline
{\large\bf LHCb collaboration}
\begin
{flushleft}
\small
R.~Aaij$^{38}$\lhcborcid{0000-0003-0533-1952},
A.S.W.~Abdelmotteleb$^{57}$\lhcborcid{0000-0001-7905-0542},
C.~Abellan~Beteta$^{51}$\lhcborcid{0009-0009-0869-6798},
F.~Abudin{\'e}n$^{57}$\lhcborcid{0000-0002-6737-3528},
T.~Ackernley$^{61}$\lhcborcid{0000-0002-5951-3498},
A. A. ~Adefisoye$^{69}$\lhcborcid{0000-0003-2448-1550},
B.~Adeva$^{47}$\lhcborcid{0000-0001-9756-3712},
M.~Adinolfi$^{55}$\lhcborcid{0000-0002-1326-1264},
P.~Adlarson$^{84}$\lhcborcid{0000-0001-6280-3851},
C.~Agapopoulou$^{14}$\lhcborcid{0000-0002-2368-0147}
% ...

{\footnotesize \it

% ...
$^{14}$Universit{\'e} Paris-Saclay, CNRS/IN2P3, IJCLab, Orsay, France\\
% ...
$^{38}$Nikhef National Institute for Subatomic Physics, Amsterdam, Netherlands\\
% ...
$^{47}$Instituto Galego de F{\'\i}sica de Altas Enerx{\'\i}as (IGFAE), Universidade de Santiago de Compostela, Santiago de Compostela, Spain\\
% ...
$^{51}$Physik-Institut, Universit{\"a}t Z{\"u}rich, Z{\"u}rich, Switzerland\\
% ...
$^{55}$H.H. Wills Physics Laboratory, University of Bristol, Bristol, United Kingdom\\
% ...
$^{57}$Department of Physics, University of Warwick, Coventry, United Kingdom\\
% ...
$^{60}$School of Physics and Astronomy, University of Glasgow, Glasgow, United Kingdom\\
$^{61}$Oliver Lodge Laboratory, University of Liverpool, Liverpool, United Kingdom\\
% ...
$^{69}$Syracuse University, Syracuse, NY, United States\\
% ...
$^{84}$Department of Physics and Astronomy, Uppsala University, Uppsala, Sweden, associated to $^{60}$\\
% ...
}
\end{flushleft}